\documentclass{aa} 
\usepackage{graphicx}
\usepackage{txfonts}
\usepackage{color}
\newcommand{\cmtwo}{cm$^{-2}$}
\newcommand{\cmthree}{cm$^{-3}$}

\newcommand{\kms}{km\,s$^{-1}$}       
\newcommand{\tas}{$T^{\star}_{\rm A}$}

\newcommand{\tex}{$T_{\rm ex}$}
\newcommand{\molh}{H$_{2}$}                              

\newcommand{\water}{H$_{2}$O}

\newcommand{\msun}{$M_{\odot}$}

\newcommand{\lapprox}{$\stackrel {<}{_{\sim}}$}
\newcommand{\about}{$\sim$}                       
\newcommand{\powten}[1]{10$^{#1}$}
\newcommand{\av}{$A_{\rm V}$}                     

\newcommand{\lte}{LTE}
\newcommand{\lsr}{LSR}

\newcommand{\amin}{$^{\prime}$}                   
\newcommand{\asec}{$^{\prime \prime}$}
\newcommand{\adeg}{$^{\circ}$}

\begin{document}
   \title{ A CS survey of small molecular cores\thanks{ Based on observations with the 12\,m NRAO (Kitt Peak, Arizona, USA), the 15\,m SEST (ESO, La Silla, Chile) and the 20\,m OSO (Onsala, Sweden) millimetre-wave antennae.}}

   \author{
   		Bengt Larsson\inst{1}
          \and
          	 Ren\'e Liseau\inst{2}       
 	}

   \institute{Stockholm University, AlbaNova University Center, Department of Astronomy, SE-106 91 Stockholm, Sweden, \email{bem@astro.su.se}
 	\and            
          	Chalmers University of Technology, Onsala Space Observatory, Department of Earth and Space Sciences, SE-439 92 Onsala, Sweden, \email{rene.liseau@chalmers.se}    		 																
              }
              
   \date{Received ; accepted}
  \abstract
   {Stars form in large clusters or in relative solitude, resulting in different mass spectra. Here, we are addressing the sites of isolated low mass star formation in the solar neighbourhood, i.e. small cloud cores within one kiloparsec.}
   {We aim at determining the physical parameters of the cores, i.e., temperature, volume density, column density and (radial) velocity fields, and the status of star formation.}
   {Surveying small dark clouds in both celestial hemispheres we study the physical conditions of low-mass star formation for detectable core masses $M\ge 0.01$\,\msun. The target list is drawn from catalogues of optically selected dark   
   cloud cores, where, the visual extinction exceeds 5 magnitudes (\av $> 5$\,mag). The selected probe is the CS molecule that needs high densities for excitation of its rotational levels. To
   gauge the state of excitation, the cores were observed in two transitions. In a limited number of cases (24), optical depths were derived from complementing lines of the rarer isotopologue C$^{34}$S for the (2-1) and (3-2) transitions.}  
   {Making small maps ($3^{\prime} \times 3^{\prime}$) maps, the 471 optically selected cores were searched for CS\,($J=2 \rightarrow 1$) and 315 (67\%) were detected (\tas\,$\ge 
   3\sigma$). Similarly, 431 cores were observed in CS\,($J=3\rightarrow 2$), of which 141 (33\%) were clearly detected (\tas $\ge 3\sigma$). In general, the position of peak CS emission does not coincide with the optically determined centre of the cores.}
   {The cores appear cold ($T<10$\,K) and, in the majority of cases, the CS emission is optically thin ($\tau<1$). On the arcminute scales of the observations, the median column density of
   carbon monosulfide is $N({\rm CS}) = 7\times 10^{12}$\,\cmtwo. For an average abundance of $N({\rm CS})/N({{\rm H}_2)}= 10^{-8}$, the median mass of the detected cores is 1.0\,\msun.
   The line shapes are most often Gaussian with widths exceeding that due to thermal broadening of \lapprox\,0.1\,\kms. The observed median FWHM\,=\,0.7\,\kms, i.e. non-thermal
   turbulence contributes dominantly to the line widths.}
   \keywords{ISM: molecules -- clouds --  evolution -- Stars: formation
           }
   \maketitle    
%

\section{Introduction}

Many cold and dense cloud cores in the interstellar medium will eventually evolve towards dynamical instability and form stars. For the cloud contraction to proceed, the generated compressional heat needs to be channeled away. This can be achieved by means of radiation losses through collisionally excited spectral lines. The first major computation of the expected cooling rates in molecular cloud  conditions were presented by Goldsmith \& Langer (1978). Water vapour (\water) was  found to be the dominant cooling agent at densities exceeding \powten{5}\,\cmthree,  irrespectively of the kinetic temperature of the gas. These early models did not consider  the effects of molecular depletion from the gas phase due to freeze-out onto dust grains: during more recent years, observations from space indicated much lower abundances than what had been assumed in the earlier works.

From an observational point of view, such studies require trustworthy statistics and, hence, large homogeneous samples. To identify the dense regions where the water cooling could be expected to become efficient, we selected molecular probes whose critical densities\footnote{The critical density is defined as the density at which radiative and collisional de-excitation rates are equal. At $T\sim 20$\,K, $n_{\rm crit} \sim 2 \times 10^5$\,\cmthree\ for CS\,(2-1) and  $1 \times 10^6$\,\cmthree\ for (3-2).} are above \powten{5}\,\cmthree. Being readily excited, the low-$J$ rotational transitions of the relatively abundant carbon monosulfide (CS) seemed a good option. 

\subsection{Sample selection}

The sample of dark cores in the Solar Neighbourhood ($d \le 1$\,kpc) is based on visual selection from the Palomar and ESO/SERC survey plates. The work by Clemens \& Barvainis (1988) yielded a large sample of isolated cores (248) in the `northern' hemisphere ($\delta \ge -33$\adeg), with equatorial coordinates accurate to 30\asec. For the `northern' catalogue, these data were complemented  with ammonia cores by Benson \& Myers (1989), DCO$^+$ cores in $\rho$\,Oph of Loren et al. (1990) and bright-rimmed globules of Sugitani et al. (1991).

Hartley et al. (1996) compiled the `Lynds' catalogue of the Southern Sky', containing 1101 entries (see also Feitzinger \& St\"uwe 1985), with equatorial coordinates better than 10\asec. This list includes Sandqvist's dark clouds (Sandqvist 1977) and Zealey's cometary globules (Zealey et al. 1983). From the catalogue by Hartley et al. (1996),  the cores corresponding to those of the `northern' sample were selected. In particular, the visual extinction 
\av\,$\ge 6$\,mag (`Density = A', which should equal Lynds' highest opacity classes = 5 and 6). The core size was initially set not to exceed 10\amin, but this effective radius was redefined to be at most 4\amin\  (see: Clemens \& Barvainis 1988). Here, the size is  defined as $r_{\rm eff} = \sqrt{a \times b}$ with the semiaxes of the source ellipse in arcminutes. This procedure could be expected to ensure a volume limited sample, since 
the number of foreground stars $N_{\rm star}$ is small and the distance $D$ to dark clouds is proportional to $N_{\rm star}$, e.g. $d = 320\,N_{\rm star}^{\,\,0.57}$\,pc over a 5\amin\ sky area (Herbst \& Sawyer 1981).  The resulting `southern' catalogue contained 193 cores.

All in all, the initial target list (`northern' + `southern') contained 538 objects. The majority of observed dense cores has a distinctly non-spherical geometry. Typical aspect ratios are of the order of 2:1 (see, e.g., Hartley et al. 1996, Clemens \& Barvainis 1988, Benson \& Myers 1989). 

\subsection{Distribution of the cores in the sky}

\begin{figure*}
  \resizebox{\hsize}{!}{\rotatebox{90}{\includegraphics{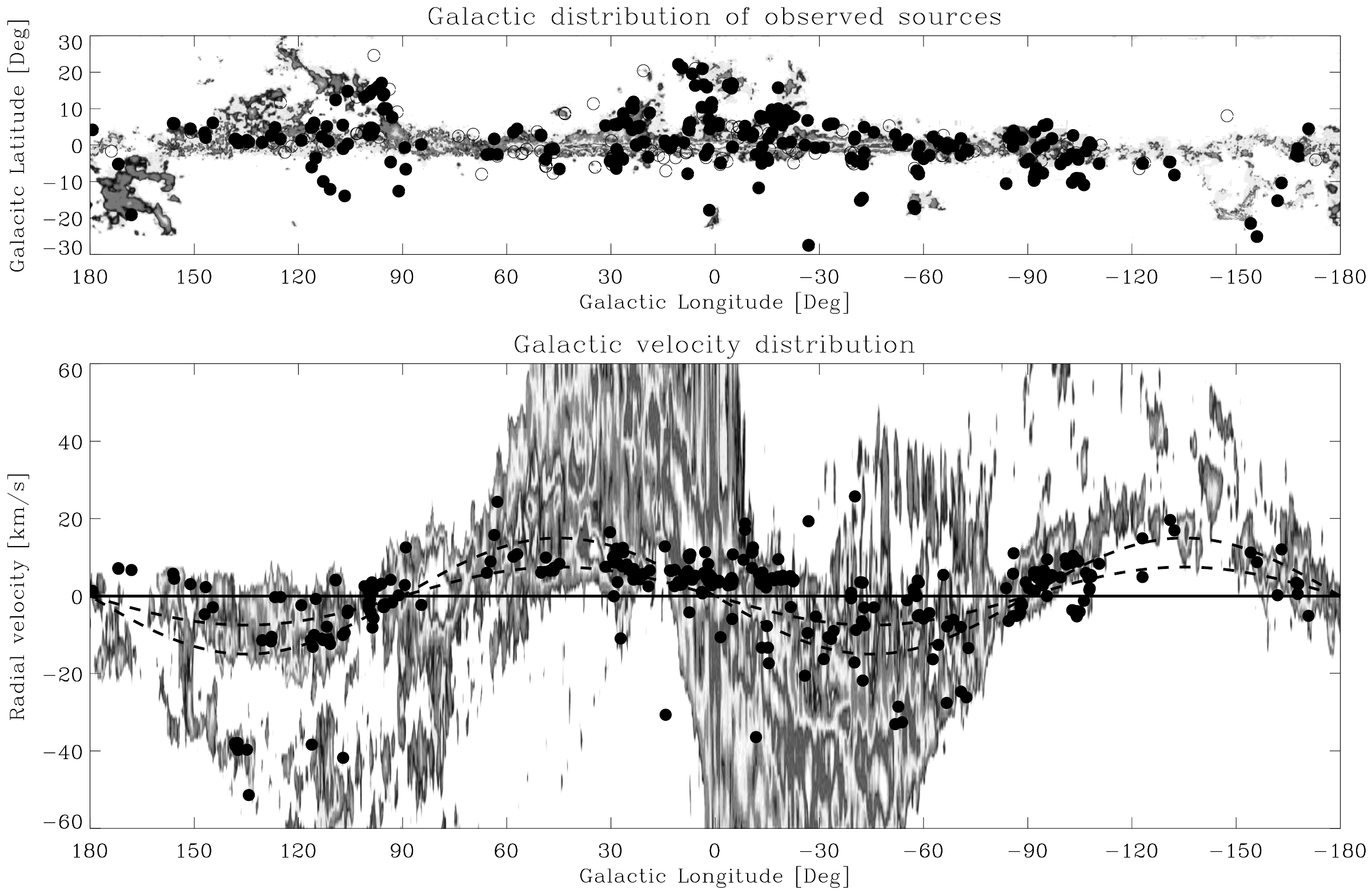}}}
  \caption{
  Small dark cloud cores are shown on maps of the large scale CO\,(1-0) distribution in the Galaxy \citep{dame1987}. {\bf Upper panel:} The sky distribution of the cores observed in rotational transitions of CS. The filled circles represent detections in CS\,(2-1), for which $T^{\star}_{\rm A} \ge 3 \sigma$. {\bf Lower panel:} The spatial \lsr-velocity distribution of these detected cores.  A simple Oort-type model is shown by the dashed lines to guide the eye, corresponding to distances of 0.5\,kpc and 1.0\,kpc, respectively. Formally, these loci are given by $\upsilon_{\rm lsr} = 13\,d\,\sin{2\,\ell}$, where $d$ is the distance in kpc and $\ell$ is the galactic longitude.}
  \label{galdist}
\end{figure*}

Not entirely unexpected, the majority  of cores is found  in, or relatively close to, the galactic plane (Fig.\,\ref{galdist}), where the high density of background stars provides a high contrast for the dark globules. Comparing the \lsr-velocities of the core gas with models of the galactic rotation indicates that most of the cores are within 1\,kpc from the Sun, i.e. that they indeed belong to the Solar Neighbourhood ($\mid\!\upsilon_{\rm lsr}\!\mid$\,\lapprox\,10\,\kms).

\subsection{Mass sensitivity of the core survey}

We adopted a limiting antenna temperature of 20\,mK (averaged over the 9-point map, cf. Sect.\,3) or 60\,mK (for any individual position) in the CS\,(2-1) line. This sensitivity corresponds roughly to a limiting \lte\ column density \footnote{This assumes kinetic temperatures in the range (10-30)\,K, line widths of \about\,0.7\,\kms, unit beam filling and a main beam efficiency of 70\%.} $N({\rm CS}) = (3 - 5) \times 10^{11}$\,\cmtwo. For a CS abundance relative to \molh, $X({\rm CS}) = 1 \times 10^{-8}$ (Irvine et al. 1987), this would correspond to a hydrogen column $N({\rm H}_2) = {\rm a\,\,few}\,\times 10^{19}$\,\cmtwo. This limit is much smaller than the expected column densities of the molecular gas associated with the high dust extinction (several \av) through the centres of the cores, since (on the average) $N({\rm H}_2) = 0.94 \times 10^{21}\,A_{\rm V}$\,\cmtwo\,mag$^{-1}$ \citep{bohlin1978}.

At the distance of 500\,pc, an object 1\amin\ in size (the approximate size of the telescope beam), i.e.  with diameter of 0.145\,pc, the minimum detectable mass would thus be of the order of \powten{-2}\,\msun. Likely effects due to subthermal excitation, line opacity, low beam filling and/or convolutedness of the object could increase this number to a higher value. However, the survey should be sensitive to the standard scale of star formation, viz. the Jeans-mass (Jeans 1927). For a of a cloud of ideal gas,
$M_{\rm Jeans} = 0.57\, \left( \frac {T}{10\,{\rm K}} \right )^{\frac {3}{2} }\,\left( \frac {n({\rm H}_2)}{10^5\,{\rm cm}^{-3}} \right )^{-\frac {1}{2} }\,M_{\odot}$. The numerical constant assumes the mean molecular weight $\mu = 2.4$ for metal rich, fully molecular gas. A cloud of mass larger than $M_{\rm Jeans}$ will be prone to gravitational collapse on a free-fall time scale
$t_{\rm ff} = 1.0 \times 10^5\, \left( \frac {n({\rm H}_2)}{10^5\,{\rm cm}^{-3}} \right )^{-\frac {1}{2} }\,{\rm yr}$. In real, rotating clouds, various pressure gradient forces, including magnetic field stresses, will potentially prolong this time scale by factors of several (e.g., Ciolek \& Mouschovias 1995). The matter will fall freely toward the centre of mass with a velocity
$\upsilon_{\rm ff} = 0.20\, \left( \frac {M}{1\,M_{\odot}} \right )^{\frac {1}{2} }\,\left( \frac {R}{0.1\,{\rm pc}} \right )^{-\frac {1}{2} }\,{\rm km\,s}^{-1}$.


\section{Observations and data reduction}

\begin{figure*}
  \resizebox{\hsize}{!}{\includegraphics{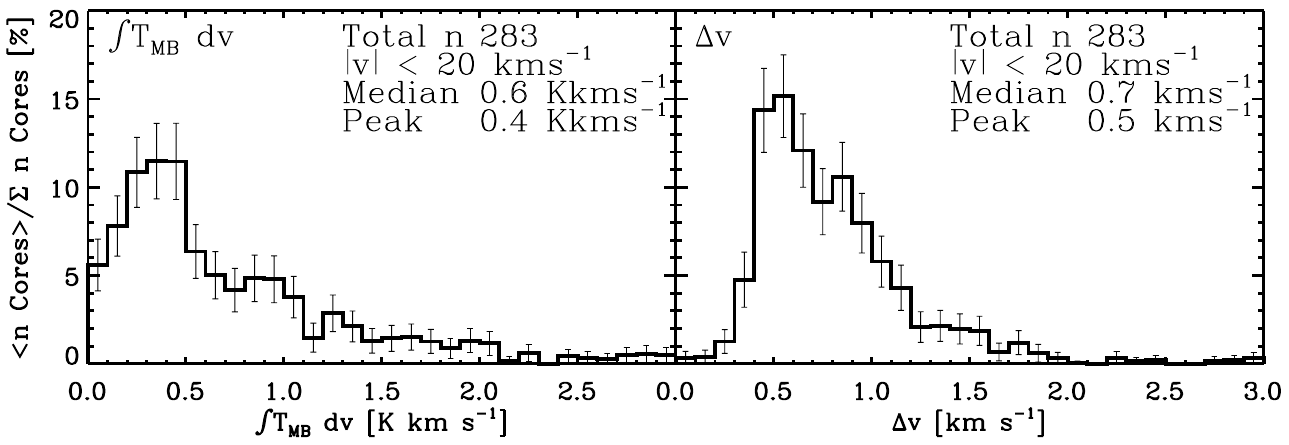}}
  \caption{{\it Left}: The distribution of integrated CS\,(2-1) line intensity  toward cores, the LSR-velocity of which fall in the interval $-20 < \upsilon_{\rm lsr} < 20$ km s$^{-1}$. {\it Right}: The CS\,(2-1) line width distribution for these cores.}
  \label{lsr21}
\end{figure*}

During several observing runs in 1996 and 1997, we exploited three different observatories, viz. the National Radio Astronomy Observatory (NRAO), Kitt Peak, Arizona, USA, the Onsala Space Observatory (OSO), Onsala, Sweden, and the European Southern Observatory, La Silla, Chile. There, we used the single-dish antennae of 12\,m, 20\,m and 15\,m diameter, respectively. The La Silla telescope is a Swedish-ESO joint facility, i.e. the Swedish ESO Submillimetre Telescope (SEST). 

The pointing accuracy of the SEST and OSO\,20\,m was regularly checked using SiO maser stars as references and was found to be about 4\asec\ (rms). The planets Jupiter and Saturn were used for the NRAO\,12\,m telescope, resulting in an accuracy of 8\asec\ (rms). From the observation of overlapping sources we determined the 3-telescope inter-calibration for the various transitions to be within 15\%--20\%. 
  
The CS\,(2-1) and (3-2), i.e. short-hand for $^{12}$C$^{32}$S\,($J\rightarrow J-1$), transitions (and their isotopic lines, see Table\,\ref{freq}) could be observed simultaneously at the SEST, whereas for the northern sources these observations were split between OSO and Kitt Peak, cf. Table\,\ref{observatories}. SIS-mixer receivers were used as front-ends at all telescopes, whereas as back-ends we used Autocorrelators, Hybrid Autocorrelators (AC, HAC) and Acousto-optical spectrometers (AOS). 

All observations were done in frequency switching mode, with switch frequencies of 4\,MHz and 6\,MHz (about 8 to 18\,\kms). Standard reductions included the removal of artefacts (interference spikes), the folding of the spectra and the polynomial fitting of the baseline and its removal. Finally, the properly weighted data were averaged to generate one spectrum per line and per position. 
 
Integration times were typically of the order of 60\,s per position. For each source, and for each line, a $3 \times 3$ map was obtained, aligned with the equatorial coordinate system and with 60\asec\ spacings. This was a trade-off between an acceptable sampling rate and a sufficient sky coverage within the available observing time. These small maps should give us a hint at the degree of compactness of the CS emission.  

\begin{table*}
  \caption{\label{freq} Rotational transitions in carbon monosulfide}
  \begin{tabular}{lcc}
    \hline
   \noalign{\smallskip}
                                & \multicolumn{2}{c}{Transition} \\
   Parameter                                                     & CS\,$(2-1)$/C$^{34}$S\,$(2-1)$ & CS\,$(3-2)$/C$^{34}$S\,$(3-2)$ \\
   \noalign{\smallskip}
    \hline
    \noalign{\smallskip}
      Tuning frequency setting $\nu_0$ [MHz]                               &  97980.968/96412.982     & 146969.049/144617.147   \\
      Wavelength of transition [mm]                                               &  3.06/3.11                & 2. 04/2.07              \\
      Energy above ground $\Delta E_{\rm u}/k$ [K]                     &  7.05 /6.94               & 14.11/13.88            \\
      Radiative transition rate$^a$ $A_{\rm ul}$ [s$^{-1}$]            &  $1.675\times 10^{-5}/1.596\times 10^{-5}$  & $6.055\times 10^{-5}/5.771\times 10^{-5}$  \\
      Collision constant$^b$ $\gamma_{\rm ul}$(20\,K) [cm$^3$\,s$^{-1}$]   &  $6.96 \times 10^{-11}$  & $4.70 \times 10^{-11}$  \\
      Critical density $n_{\rm crit}$(20\,K) [\cmthree ]                   &  $2.4 \times 10^{5}$     & $1.3 \times 10^{6}$     \\
   \noalign{\smallskip}
    \hline
   \noalign{\smallskip}    
  \end{tabular}

$^a$ see also: Pineiro et al. (1987), Chandra et al. (1995). \\
$^b$ Turner et al. (1992).
 \end{table*}

\begin{table*}
  \caption{\label{observatories} Millimetre wave observatories}
  \begin{tabular}{lcccc}
    \hline
   \noalign{\smallskip}
                                & \multicolumn{4}{c}{Telescope/Transition} \\
                                &  20\,m OSO & 12\,m NRAO & 15\,m SEST & 15\,m SEST   \\
      Parameter                 & CS\,$(2-1)$ & $(3-2)$ & $(2-1)$  &  $(3-2)$ \\
   \noalign{\smallskip}
    \hline
    \noalign{\smallskip}
      Observing run   &  Jan\,97, Mar\,97 &  May\,97, Jun\,97 & Oct\,96, Jun\,97 &  Oct\,96, Jun\,97 \\
      Beam size (HPBW) [\arcsec]          &      37 &      41 &     50 &    33 \\
      Main beam efficency $\eta_{\rm mb}$ &    0.56 &    0.55 &   0.70 &  0.66 \\
      $T_{\rm sys}$ [K]                   &     390 &     270 &    160 &   190 \\
      Frequency Throw [MHz]               &       6 &       4 &      4 &     4 \\
      Spectrometer$^{a}$                  &     AC  &     HAC &  AOS   &   AOS \\
      Bandwidth [MHz]                     &      80 &    37.5 &     43 &    43 \\
      Number of channels                  &    1600 &     768 &   1000 &  1000 \\
      Resolution, $\Delta \upsilon$ [km s$^{-1}$]&    0.31 &    0.20 &   0.26 &  0.18 \\
      $T^{\star}_{\rm A}$ (rms) [mK channel$^{-1}$] & 89 & 67 &     68 &    79 \\
   \noalign{\smallskip}
    \hline
   \noalign{\smallskip}    
 \end{tabular}

$^{a}$ AC = AutoCorrelator, HAC = Hybrid AutoCorrelator, AOS = AcoustoOptical Spectrometer.
 \end{table*}

\section{Results}

\subsection{CS\,(2-1) and CS\,(3-2) detection rates}

Of the originally selected sources, 471 (88\%) were observed in CS\,(2-1) and 431 (80\%) in CS\,(3-2). Table\,\ref{tab_obs_cs} summarises the observational results. Detections at the $\ge 3\sigma$ level of the antenna temperature \tas\ were two thirds (315) in the (2-1) and one third (141) in the (3-2) transition. These relatively high detection rates clearly support the view that the selected dark dust patches in the sky are sites of high molecular gas density, approaching those of the critical values of the CS transitions.

\subsection{Line intensities and line widths}

\begin{figure}
  \resizebox{\hsize}{!}{\includegraphics{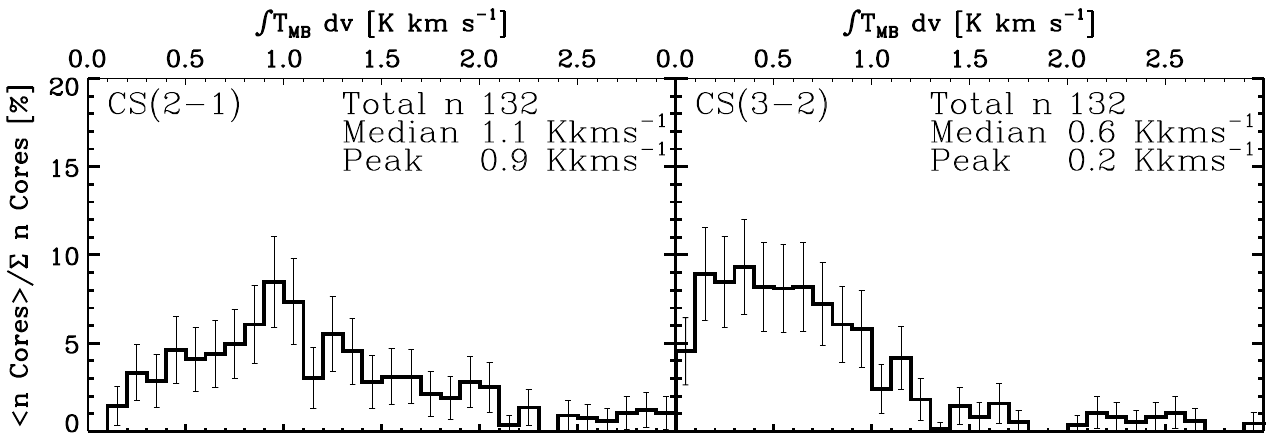}}
  \resizebox{\hsize}{!}{\includegraphics{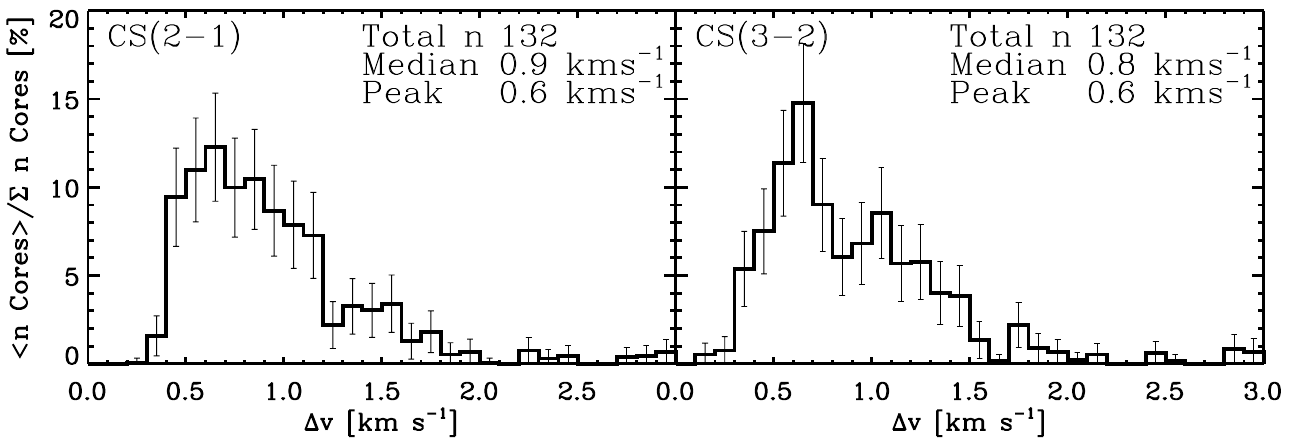}}
  \caption{Same as in Fig.\,\ref{lsr21} but for a reduced sample where both CS\,(2-1) and CS\,(3-2) lines were detected and  for which the LSR- velocity $-20 < \upsilon_{\rm lsr} < 20$ km s$^{-1}$. {\it Left}: CS\,(2-1) and {\it Right}: CS\,(3-2).}
  \label{lsr32}
\end{figure}

Limiting the sample to cores with $\mid \!\!\upsilon_{\rm LSR}\!\!\mid < 20$\,\kms\ results in the distributions\footnote{Note regarding all distribution plots: Each bin is the average of 10 different samplings of the data. The bin size is kept constant, but the mid point of the bin is shifted by 10\% of the bin size for each of the 10 samplings. The error bars represent $\pm \sqrt{\sigma^{2}_{S} + \sigma^{2}_{N}}$ where $\sigma_{S}$ is the mean error of that average and $\sigma_{N}$ is the square root of the number of cores in that bin.} for 283 CS\,(2-1) sources shown in Fig.\,\ref{lsr21}. The median line intensity is $\int \!T_{\rm mb} d\upsilon = 0.6$\,K\,\kms\ and the observed median line width is ${\rm FWHM} = 0.7$\,\kms. That this result is not limited by the resolution of the spectrometers is verified in Fig.\,\ref{lsr32}, where similar distributions are shown for objects for which both valid (\tas\,$>3\sigma$) CS\,(2-1) and CS\,(3-2) data are available (132 sources), since the higher frequency lines provide a better velocity resolution (cf. Table\,\ref{observatories}).

These observed line widths are much larger than those from purely thermal broadening, viz. $\upsilon_{\rm th} = 0.06\,\sqrt{T/(10\,{\rm K})}$\,\,\kms, and are thus dominated by other types of motion, including non-thermal turbulence. 

\subsection{Positional coincidences and source sizes}

As described in the previous section, 9-point maps ($3^{\prime} \times 3^{\prime}$) were obtained for each core. Inspection of these small maps (Table\,\ref{tab_obs_cs}) reveals that the CS\,(2-1) peak emission most often does not coincide with the optically determined centre position. The fraction of sources for which positional agreement exists amounts to 38\%. 
One explanation could be that the molecular line observations better probe the regions of maximum density (core centre) than the optical data which saturate at \av$\sim 5$\,mag, i.e., becoming indiscriminant for higher opacity regions.


\begin{figure}
  \resizebox{\hsize}{!}{\includegraphics{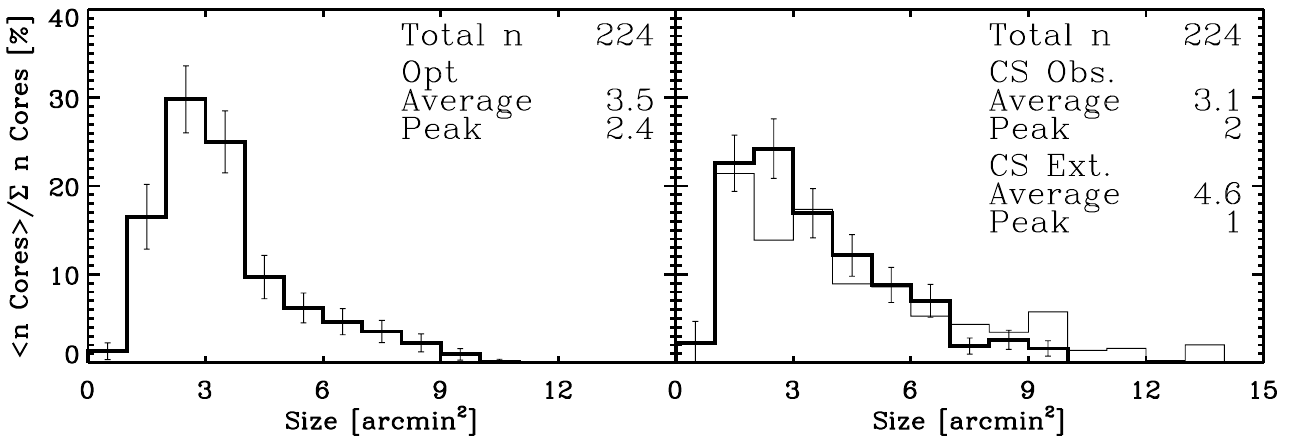}}
  \caption{The source size distribution. {\it Left panel}: The optically estimated size (from the source catalogues). {\it Right panel}: The size from the CS\,(2-1) observations, where the observed maps are represented by the thick line, whereas the thin line depicts the extended distribution (see the text).}
  \label{size}
\end{figure}

In Fig.\,\ref{size}, the optically determined core sizes and those found from the CS\,(2-1) observations are shown. The selection is biased and the CS mapping incomplete. However, from the figure it is evident that the sample is
peaked towards small CS cores with an average value of 3.1\,arcmin$^2$. Even though the observed map is small and under-sampled, one can still get a rough estimate of the core extent in CS, simply by adding the number of map positions, each 1\,arcmin$^2$ in size, where the detected signal is larger than half the peak intensity.

\begin{figure}
  \resizebox{\hsize}{!}{\includegraphics{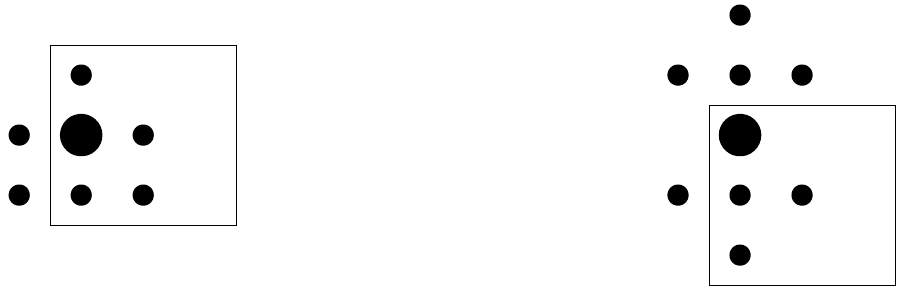}}
  \caption{Examples of mirroring the observed maps of extended sources (large dots symbolise peak integrated line emission, whereas the smaller symbols refer to those  positions where the emission is above 50\% of the peak value).   {\it Left}: The peak is at a map edge. {\it Right}: The peak is in a map corner.}
  \label{sizemirror}
\end{figure}

As already mentioned, the peak intensity of the CS line is not always found at the centre position. In addition, several cores show evidence of being extended on scales larger than our limited maps. We will therefore potentially underestimate the sizes of a number of cores. By assuming that the core is symmetric around its peak and by mirroring the map with respect to this position, a somewhat better estimate of the source size can be found (see Fig.\,\ref{sizemirror} for examples).

Finally, the resulting size distribution is also shown in Fig.\,\ref{size}. The average core size is about $(4 \pm 1)$\,arcmin$^2$ (optical: $3.5 \pm 0.2$, CS\,(2-1): $3.1 \pm 0.2$, extrapolated CS: $4.6 \pm 0.3$). These independent size estimates (optical and CS) are in good agreement, indicative of a tight relation between the dust and the molecular gas (the average size ratio $\log { [{\rm Optical/CS\,(2-1)}] } = -0.085$, with a variance of 0.092). This indicates that, on the average, the beam should to a large extent be filled around the peak position.

\subsection{The volume limited sample}

The number of cores $N$ subtending a given solid angle $\Omega$ can be written as  $\frac{dN}{d\Omega} = \frac{dN}{dD}\frac{dD}{d\Omega}$, where $D$ is the distance to the core. For the case that the cores are of the same size and are evenly distributed in space, the following applies  $\frac{dN}{dD}\propto \Omega^{-1}$ and that $\frac{dD}{d\Omega} \propto \Omega^{-3/2}$. This integrates to $N(\Omega) \propto \Omega^{-3/2}$, so that the slope of this distribution is $\frac {d\log N} {d\log \Omega} = -1.5$.

The core size distributions of Fig.\,\ref{size} all have slopes of about $-1.5$, consistent with the result of Eq.\,12 (optical: $-1.62 \pm 0.4$, CS\,(2-1): $-1.26 \pm 0.28$, extrapolated CS: $-1.28 \pm 0.16$). We take this to support our view that the selected core sample is reasonably homogeneous and volume limited.

\section{Discussion}

\subsection{Optically thin emission: C$^{34}$S\,(2-1) and C$^{34}$S\,(3-2)}

The solution to the equation of radiative transfer of molecular radio lines reads $T = f_{\rm beam}\, \left [\mathcal{F}(T_{\rm ex}) - \mathcal{F}(T_{\rm bg}) \right ] \,\left ( 1 - e^{-\tau} \right )$, where $\mathcal{F}(T) = \frac {h\,\nu/k} {e^{h\,\nu/kT} - 1}$ and the symbols have their usual meaning. $f_{\rm beam}$ is the beam filling factor, $T_{\rm ex}$ is the excitation temperature of the transition (the line source function), $T_{\rm bg}$ represents a background radiation field, which at mm-wavelengths is the cosmic microwave radiation ($T_{\rm bg} = 2.725$\,K), and $\tau$ is the line optical depth.

Assuming that the lines of C$^{34}$S are formed in the same layers as the corresponding CS lines (= C$^{32}$S) and that the line opacities scale as their relative abundances, i.e. $x = \tau({\rm C^{32}S})/\tau({\rm C^{34}S)} = X{\rm (^{32}S)/X(^{34}S)}\, (\,= 22.5)$, means that the radiative transfer equation for the ratio of these lines reduces to $ \frac {1 - e^{- x\,\tau(34)}} {1 - e^{- \tau(34)}} \approx \frac {T_{\rm mb}{\rm (CS)}_{(J,\,J-1)}} {T_{\rm mb}{\rm (C^{34}S)}_{(J,\,J-1)}}$. This transcendental equation can be solved iteratively for the optical depth in the C$^{34}$S lines,  $\tau(34)$, and these are presented for a number of cores in Table\,\ref{34S}. Quite in accordance with expectation, these values are all smaller than unity.

For the optically thin C$^{34}$S lines we  derive the excitation temperature and column density of the molecule from degenerated rotation diagrams (see, e.g., Goldsmith \& Langer 1999).
For these two transitions, i.e. the C$^{34}$S\,(2-1) and (3-2) lines, the excitation temperatures, $T_{\rm ex}$, and column densities, $N{\rm (C^{34}S)}$, are listed in Table\,\ref{34S}. Having determined the opacities from the C$^{34}$S lines, the same analysis can be done for the CS lines, which results in the values presented in Table\,\ref{32S}\,\footnote{Core IDs and coordinates are given in Table\,6}. From this it is evident that the limited number of cores with detectable C$^{34}$S emission has only moderately high optical depths in the lines of the main isotope of sulfur. The majority of cores can be expected therefore to exhibit $\tau$\,\lapprox\,1.

\begin{table*}
  \caption{\label{34S} Excitation temperatures and column densities from C$^{34}$S\,(2-1) and (3-2) observations}
  \begin{tabular}{rrrcrrrcrrr}
  \hline
   \noalign{\smallskip}  
      & \multicolumn{4}{c}{C$^{34}$S\,(2--1)} & \multicolumn{4}{c}{C$^{34}$S\,(3--2)} &  &  \\
 Core & $T_{\rm mb}$ & $\Delta T$  & $\int \!T_{\rm mb}\,dv$ & $\tau_{21}$
      & $T_{\rm mb}$ & $\Delta T$  & $\int \!T_{\rm mb}\,dv$ & $\tau_{32}$
                                                                                & $T_{\rm ex}$ & $N{\rm (C^{34}S)}$ \\ 
      & (K)          &      (K)    & (K\,\kms)               &
      & (K)          &      (K)    & (K\,\kms)               &     
                                                                                & (K)          & (\cmtwo) \\                                                               
    \noalign{\smallskip}  
  \hline
    \noalign{\smallskip}  
   40 &   0.32 &   0.04 &   0.36 &    0.14 &   0.19 &   0.03 &   0.09 &    0.15 &   3.7  &  $9.4 \times 10^{12}$ \\
   41 &   0.11 &   0.03 &   0.17 &    0.07 &        &   0.02 &        & $<$0.10 &        &           \\
   52 &   0.21 &   0.04 &   0.29 &    0.14 &   0.12 &   0.02 &        &    0.08 &        &           \\
   58 &   0.14 &   0.03 &   0.03 &    0.17 &        &   0.02 &        & $<$0.15 &        &           \\
   60 &   0.13 &   0.03 &   0.06 &    0.09 &        &   0.02 &        & $<$0.12 &        &           \\
   61 &   0.15 &   0.03 &   0.09 &    0.12 &        &   0.02 &        & $<$0.10 &        &           \\
   64 &   0.13 &   0.03 &   0.15 &    0.09 &   0.08 &   0.02 &   0.04 &    0.16 &   4.0  &  $3.3 \times 10^{12}$ \\
   65 &   0.12 &   0.03 &   0.08 &    0.08 &   0.09 &   0.02 &   0.08 &    0.16 &  12.6  &  $6.7 \times 10^{11}$ \\
   66 &   0.14 &   0.03 &   0.14 &    0.08 &        &   0.02 &        & $<$0.08 &        &           \\
   74 &   0.21 &   0.03 &   0.10 &    0.16 &   0.09 &   0.02 &   0.06 &    0.04 &   7.6  &  $8.4 \times 10^{11}$ \\
   76 &   0.21 &   0.04 &   0.15 &    0.10 &   0.08 &   0.02 &   0.07 &    0.03 &   5.8  &  $1.6 \times 10^{12}$ \\
   82 &        &   0.04 &        & $<$0.22 &        &   0.02 &        & $<$0.30 &        &           \\
   88 &   0.21 &   0.03 &   0.09 &    0.17 &   0.09 &   0.02 &   0.04 &    0.13 &   5.8  &  $9.8 \times 10^{11}$ \\
   90 &   0.15 &   0.04 &   0.13 &    0.02 &   0.13 &   0.03 &   0.04 &    0.09 &   4.5  &  $2.0 \times 10^{12}$ \\
   92 &   0.19 &   0.02 &   0.16 &    0.12 &        &   0.01 &        & $<$0.03 &        &           \\
  100 &   0.16 &   0.04 &   0.08 &    0.18 &        &   0.03 &        & $<$0.14 &        &           \\
  102 &        &   0.03 &        & $<$0.15 &        &   0.02 &        & $<$0.15 &        &           \\
  106 &   0.09 &   0.03 &   0.08 &    0.13 &   0.10 &   0.02 &   0.05 &    0.29 &   7.1  &  $7.4 \times 10^{11}$ \\
  115 &        &   0.02 &        & $<$0.04 &        &   0.01 &        & $<$0.10 &        &           \\
  117 &   0.13 &   0.03 &   0.11 &    0.09 &   0.08 &   0.02 &   0.06 &    0.14 &   6.9  &  $9.7 \times 10^{11}$ \\
  132 &   0.14 &   0.06 &   0.21 & $<$0.13 &        &   0.04 &        & $<$0.14 &        &           \\
    \noalign{\smallskip}  
  \hline
  \end{tabular}
\end{table*}

\begin{table*}
  \caption{\label{32S} Same as in Table\,\ref{34S}, but for the corresponding C$^{32}$S\,(2-1) and (3-2) observations}
  \begin{tabular}{rrrcrrrcrrr}
  \hline
   \noalign{\smallskip} 
      & \multicolumn{4}{c}{CS\,(2--1)} & \multicolumn{4}{c}{CS\,(3--2)} &  &  \\
 Core & $T_{\rm mb}$ & $\Delta T$  & $\int \!T_{\rm mb}\,dv$ & $\tau_{21}$
      & $T_{\rm mb}$ & $\Delta T$  & $\int \!T_{\rm mb}\,dv$ & $\tau_{32}$
                                                                                & $T_{\rm ex}$  & $N$(CS) \\ 
                                                                                
      & (K)          &      (K)    & (K\,\kms)               &
      & (K)          &      (K)    & (K\,\kms)               &     
                                                                                & (K)          & (\cmtwo) \\        
   \noalign{\smallskip}   
    \hline    
    \noalign{\smallskip} 
   40 &   2.33 &   0.08 &   1.78 &    3.23 &   1.27 &   0.07 &   1.04 &    3.43 &   6.9  &  $1.6 \times 10^{13}$ \\
   41 &   1.31 &   0.05 &   1.37 &    1.58 &   0.62 &   0.06 &   0.67 & $<$2.32 &   5.9  &  $1.4 \times 10^{13}$ \\
   52 &   1.62 &   0.05 &   3.92 &    3.04 &   1.30 &   0.12 &   3.00 &    1.71 &   9.5  &  $3.1 \times 10^{13}$ \\
   58 &   0.85 &   0.07 &   0.94 &    3.88 &   0.45 &   0.08 &   0.69 & $<$3.37 &   9.0  &  $7.6 \times 10^{12}$ \\
   60 &   1.28 &   0.05 &   1.10 &    2.10 &   0.57 &   0.08 &   0.44 & $<$2.60 &   5.1  &  $1.4 \times 10^{13}$ \\
   61 &   1.27 &   0.05 &   1.33 &    2.61 &   0.62 &   0.05 &   0.71 & $<$2.35 &   6.5  &  $1.3 \times 10^{13}$ \\
   64 &   1.27 &   0.06 &   2.09 &    2.05 &   0.57 &   0.06 &   0.77 &    3.52 &   4.8  &  $2.9 \times 10^{13}$ \\
   65 &   1.34 &   0.06 &   1.51 &    1.70 &   0.60 &   0.07 &   0.55 &    3.59 &   4.7  &  $2.1 \times 10^{13}$ \\
   66 &   1.47 &   0.06 &   1.53 &    1.84 &   0.60 &   0.06 &   0.70 & $<$1.87 &   5.6  &  $1.7 \times 10^{13}$ \\
   74 &   1.38 &   0.10 &   0.93 &    3.56 &   1.42 &   0.12 &   0.65 &    0.84 &   8.4  &  $7.6 \times 10^{12}$ \\
   76 &   1.95 &   0.06 &   1.92 &    2.24 &   1.27 &   0.06 &   1.26 &    0.69 &   7.9  &  $1.6 \times 10^{13}$ \\
   82 &   0.55 &   0.07 &   0.69 & $<$4.85 &   0.27 &   0.08 &   0.32 & $<$6.68 &   5.6  &  $7.6 \times 10^{12}$ \\
   88 &   1.28 &   0.11 &   1.10 &    3.91 &   0.74 &   0.12 &   0.52 &    2.91 &   5.8  &  $1.2 \times 10^{13}$ \\
   90 &   2.60 &   0.06 &   2.64 &    0.52 &   1.30 &   0.07 &   1.50 &    2.04 &   6.8  &  $2.4 \times 10^{13}$ \\
   92 &   1.58 &   0.06 &   1.51 &    2.74 &   0.65 &   0.07 &   0.63 & $<$0.72 &   5.2  &  $1.8 \times 10^{13}$ \\
  100 &   0.92 &   0.10 &   0.31 &    4.06 &   0.60 &   0.13 &   0.18 & $<$3.10 &   6.8  &  $2.9 \times 10^{12}$ \\
  102 &   0.56 &   0.07 &   0.48 & $<$3.45 &   0.37 &   0.09 &   0.18 & $<$3.43 &   4.9  &  $6.4 \times 10^{12}$ \\
  106 &   0.70 &   0.07 &   0.98 &    2.95 &   0.41 &   0.09 &   0.37 &    6.62 &   4.8  &  $1.3 \times 10^{13}$ \\
  115 &   0.70 &   0.09 &   0.80 & $<$0.94 &   0.41 &   0.11 &   0.22 & $<$2.20 &   4.0  &  $1.6 \times 10^{13}$ \\
  117 &   1.32 &   0.07 &   1.39 &    2.11 &   0.58 &   0.08 &   0.63 &    3.19 &   5.6  &  $1.5 \times 10^{13}$ \\
  132 &   1.40 &   0.11 &   2.08 & $<$2.96 &   0.90 &   0.11 &   0.52 & $<$3.05 &   3.8  &  $5.1 \times 10^{13}$ \\
    \noalign{\smallskip}   
    \hline
  \end{tabular}
\end{table*}

From Tables\,\ref{34S} and \ref{32S} it is evident that the derived excitation temperatures are remarkably low. If these were meant to represent true gas kinetic temperatures, we have to conclude that assuming LTE (Local Thermodynamic Equilibrium) would likely be erroneous. Else, for unit beam filling, it would seem that the volume densities of cores, averaged over the beam, are generally lower than the critical densities of the transitions.

\subsection{CS column densities and core masses}

\begin{figure}
  \resizebox{\hsize}{!}{\includegraphics{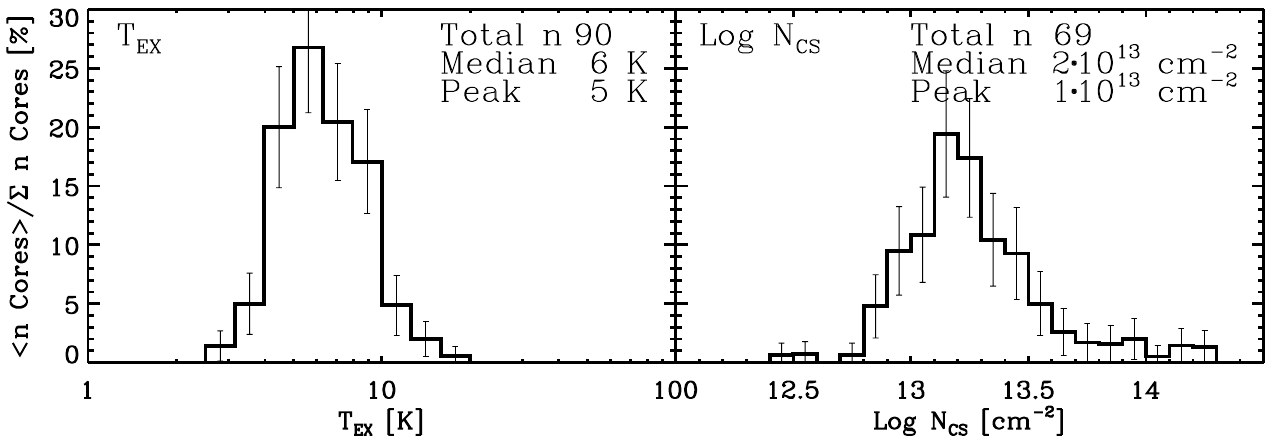}}
  \caption{{\it Left}: Excitation temperatures derived from the two transitions of CS\,(2-1) and CS\,(3-2), under the assumption of low line opacities. {\it Right}: The CS column densities derived from these transitions. For both distributions, the parameters are larger than $3\sigma$, resulting in different total numbers.}
  \label{tex21_32}
\end{figure}

For the present sample, the analysis of the previous section, including maps of two transitions and of their weak isotopic lines, could be done only for a limited number of cores. The size of the analysable sample can be increased by applying the C$^{34}$S method to the set of CS data for which two transitions have been observed. Assuming optically thin emission in the (2-1) and (3-2) lines of the main isotope (justified shortly),  results in the $T_{\rm ex}$ distribution shown in Fig.\,\ref{tex21_32} for 90 cores (\tex\,$>3\sigma$). These temperatures are, again, generally quite low, 6\,K on the average. For 69 cores ($N>3\sigma$), the data quality permits the determination of the CS column densities, which are of the order of \powten{13}\,\cmtwo\ and thus consistent with the values of Table\,\ref{32S}.

\begin{figure}
  \resizebox{\hsize}{!}{
                        \includegraphics{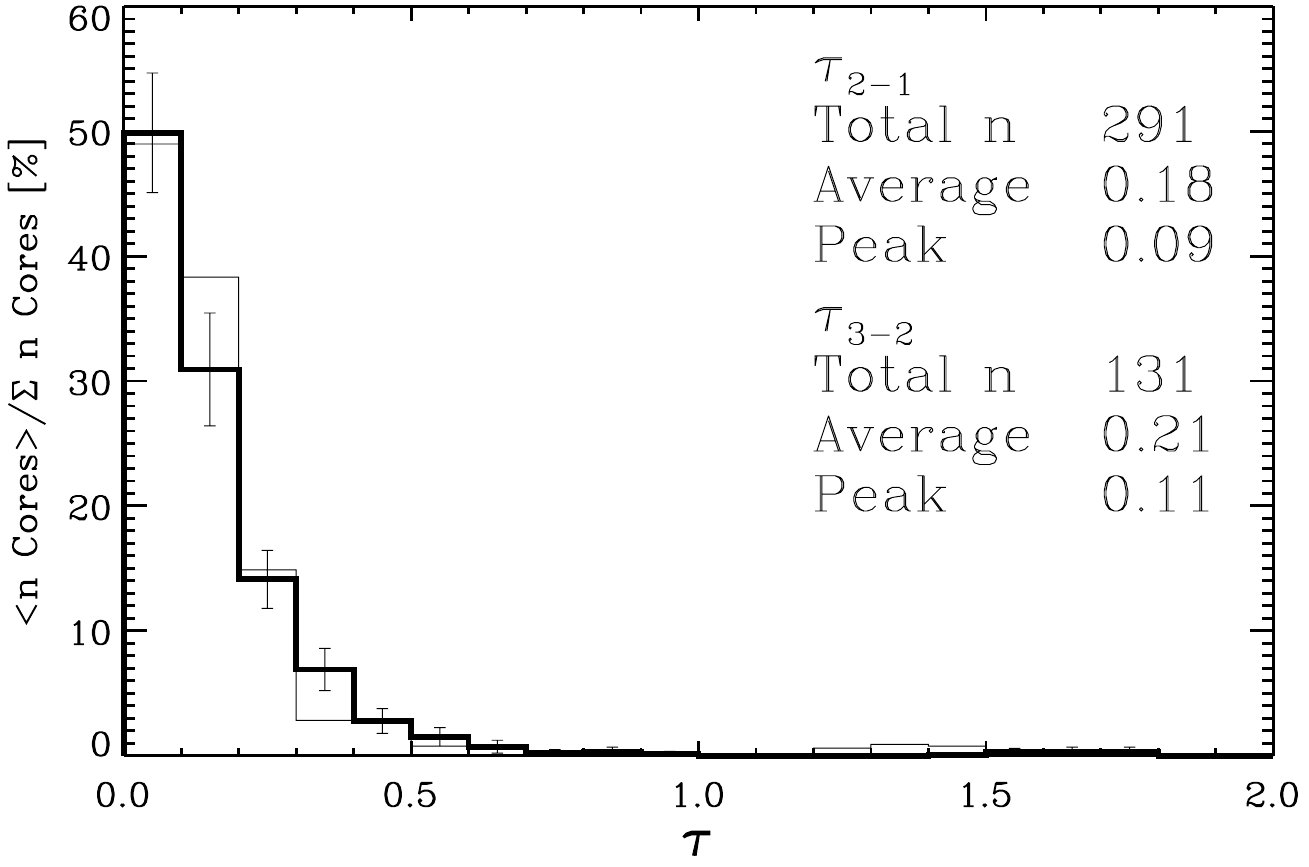}
                       }
  \caption{Distribution of the line optical depth $\tau$ in the CS\,(2-1) transition (thick line) for a total of 291 cores and in the CS\,(3-2) transition (thin line) under the assumption of a uniform excitation temperature of 10\,K (see the text).}
  \label{tau10K}
\end{figure}

These column densities are not extremely sensitive to the actual excitation temperature (within reasonable limits, of course), so that the adoption of a reasonable average value would permit us to use the entire data set. Consequently, we used (somewhat arbitrarily) the value of 10\,K \citep[see, e.g.,][]{benson1989}. However, as will be demonstrated a posteriori, this choice is not critical to our final results (see below). Once $T_{\rm ex}$ is known, the line optical depth $\tau$ can be determined from the observations. This is shown in Fig.\,\ref{tau10K}, where the distributions of both the CS\,(2-1) (291 cores) and (3-2) (131 cores) line depths are depicted. As expected, the optical depth in the (3-2) transition is larger than that in the (2-1) line. Also evident from the figure is the fact that the majority of cores is optically thin in these lines. 

\begin{figure}
  \resizebox{\hsize}{!}{
                        \includegraphics{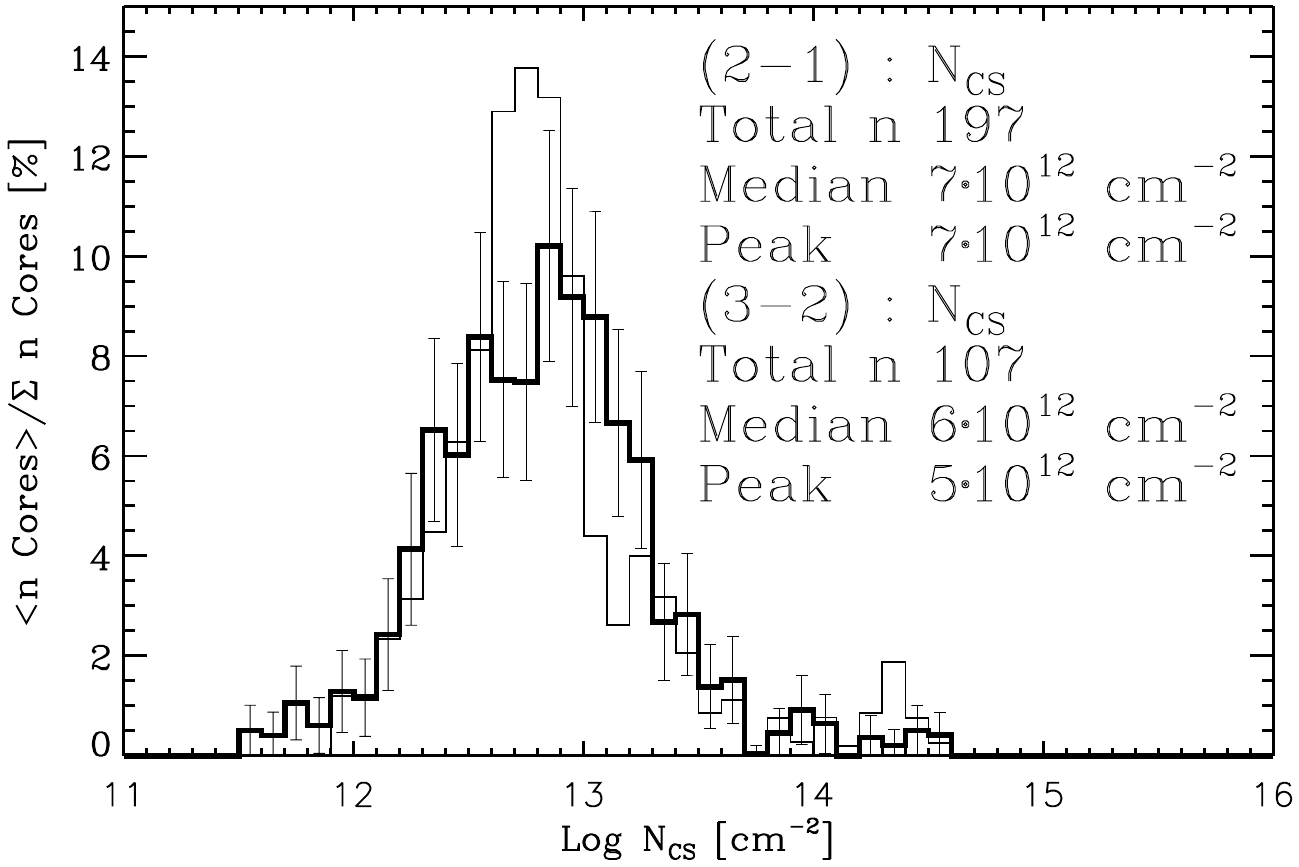}
                       }
  \caption{CS column densities for $T_{\rm kin}=10$\,K derived from the CS\,(2-1) (thick line) and CS\,(3-2) (thin line) transitions. These data are consistent with $N{\rm (CS)} = 7 \times 10^{12}$\,\cmtwo\ for the majority of cores.}
  \label{column10K}
\end{figure}

From these optically thin transitions we derive the CS column densities. These are shown in Fig.\,\ref{column10K}, separately for the CS\,(2-1) and (3-2) observations. Obviously, these distributions are very similar, the (2-1) data being less noisy. Both yield practically identical results, i.e. the median CS column density of nearly 200 cores is $N({\rm CS}) = 7 \times 10^{12}$\,\cmtwo. This is comparable with the values of Table\,\ref{32S} which were based on a more detailed and accurate analysis of the strong line cores. 

\begin{figure}
  \resizebox{\hsize}{!}{
                        \includegraphics{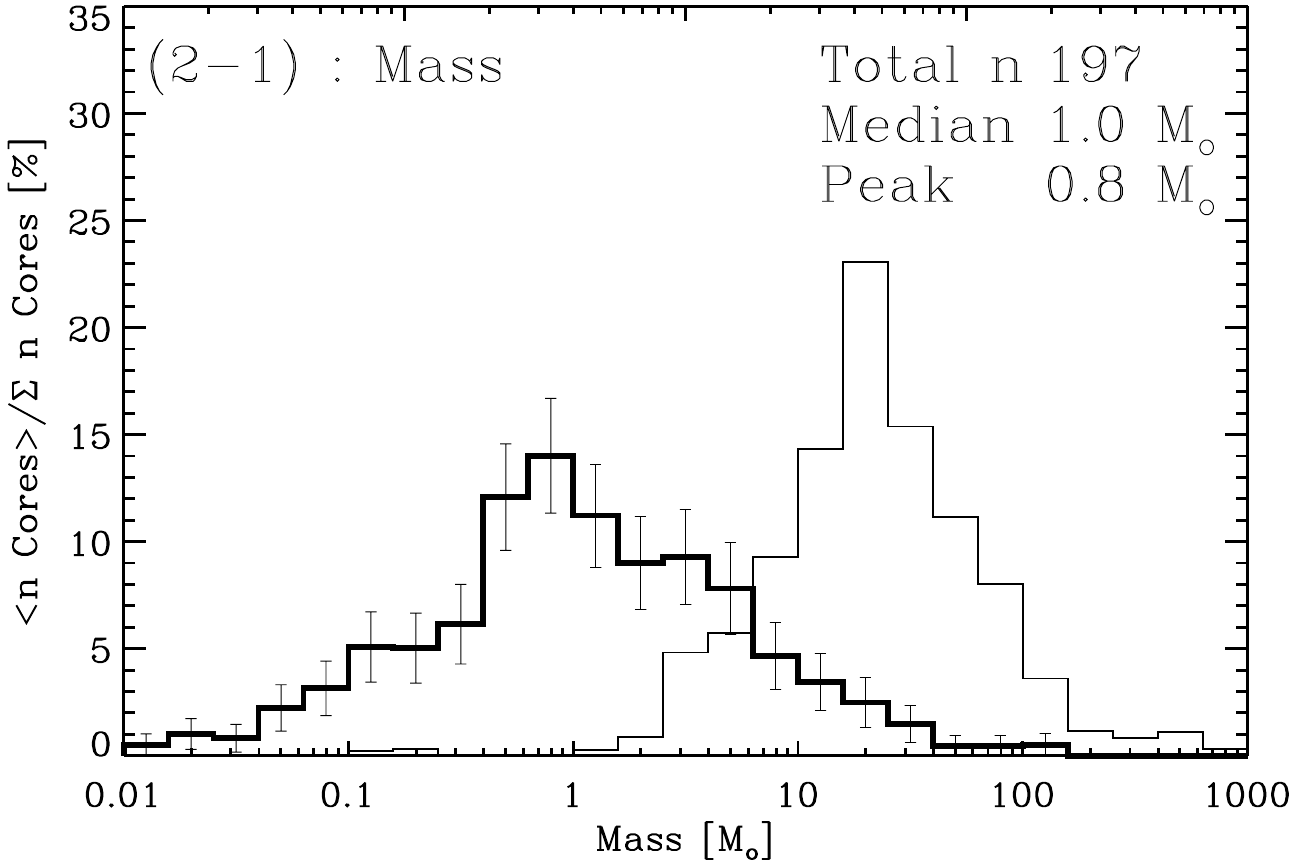}
                       }
  \caption{The mass distribution for 197 cores from CS\,(2-1) observations is shown by the thick line. The assumptions include $T_{\rm kin} = 10$\,K, $X({\rm CS}) = 10^{-8}$ and $d = 500$\,pc yielding the median mass of 1\,\msun. The light-line histogram shows the `virial' mass of these cores, based on their observed line widths.} 
  \label{mass10K}
\end{figure}

Assuming a `universal' CS abundance relative to \molh\ of $X({\rm CS}) = 10^{-8}$ and a common distance of 500\,pc, the column density distribution of Fig.\,\ref{column10K} can be transformed into a core mass distribution (Fig.\,\ref{mass10K}). As before (Sect.\,4.3), this procedure is based on (for the area) adding the pixels with valid data ($>3\sigma$) above the 50\% peak value. For the sample of nearly 200 cores, the observed apparent mass range is from \powten{-2}\,\msun\ to \powten{2}\,\msun, where the low mass cut-off is set by the lack of observational sensitivity. The median mass is 1\,\msun, indicating that perhaps up to about \powten{2} cores are Jeans-unstable. 

\begin{figure}
  \resizebox{\hsize}{!}{
                        \includegraphics{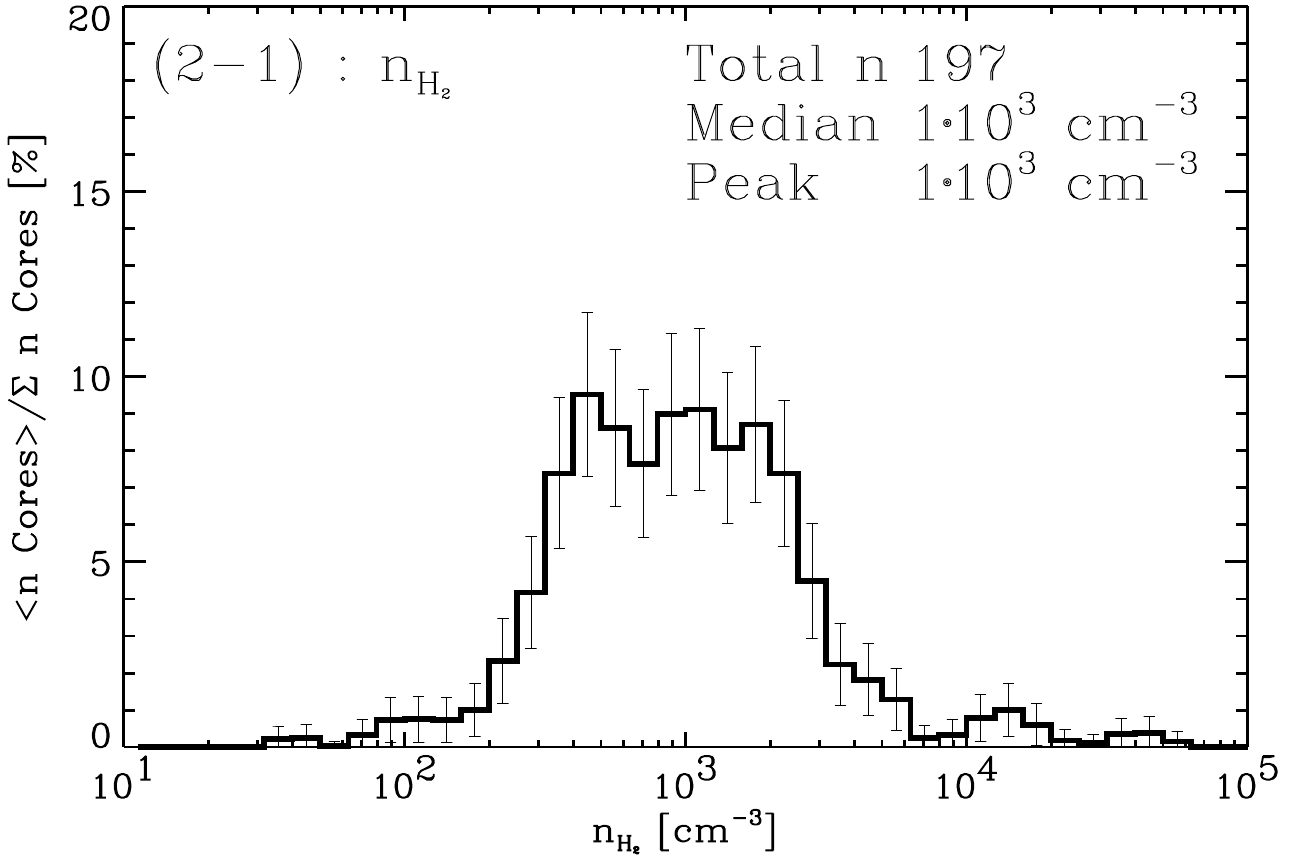}
                       }
  \caption{Volume density distribution $n$(\molh) for 197 cores from CS\,(2-1) observations (same assumptions as in caption of Fig.\,\ref{mass10K}).} 
  \label{dens10K}
\end{figure}

The corresponding distribution of volume densities $n$(\molh) can be obtained assuming that the extent of the core along the line of sight is similar to that in the plane of the sky, i.e. $n \approx N/\sqrt{\pi \times {\rm area}}$ (Fig.\,\ref{dens10K}). We find that the median density is $n({\rm H}_2) = 10^3$\,\cmthree, at which the observed median line width (FWHM\,=\,0.7\,\kms) would correspond to a magnetic field strength of $B = 16\,\mu$G, if the CS lines were dominated by magnetic broadening. The corresponding critical mass would be $M_{\rm crit} = 1.25\,(B/15\,\mu{\rm G})\,(R/0.1\,{\rm pc})^2$\,\msun\ (Mouschovias \& Spitzer 1976). The results of Fig.\,\ref{mass10K} could then suggest that less than half of the cores are potentially unstable.

However, from the results shown in Fig.\,\ref{dens10K}, it is also evident that the (averaged) densities are much lower than the critical density of the CS\,(2-1) transition. This is clearly inconsistent with the assumption of LTE, since thermalisation of the level populations requires densities at least one order of magnitude above $n_{\rm crit}$. 

The actual value of the CS abundance could be lower than that assumed above. This is also suggested by the large relative offset of the virial mass distribution shown in Fig.\,\ref{mass10K}, which is based on the observed line widths of the CS\,(2-1) line and the common distance of 500\,pc. In fact, large variations for different locations in the Galaxy are indicated in the tabulation by Irvine et al. (1987). In particular, for the two well studied dark clouds TMC\,1 and L\,134N, $X$(CS) differs by one order of magnitude, with \powten{-9} in L\,134N, possibly due to chemical evolution (Ohishi et al. 1992). If evolutionary effects are important for CS, then the application of a single abundance value to an ensemble of sources would of course be meaningless. On the other hand, this would open up the opportunity to use CS (in combination with other species) as a chemical clock. 

\subsection{Line profiles}

In order to avoid spurious effects due to the noise in the observed spectra we selected those observations the line intensity of which is above the $7\sigma$ level of the rms-noise (this particular value $T_{\rm mb,\,0} > 7\sigma$ was found empirically). Ideally, the velocity of the object with respect to the telescope $\upsilon_0$ should be determined from optically thin lines, that are known to be symmetric. However, since we have only a limited amount of C$^{34}$S data, the central (= systemic) velocities of the CS\,(2-1) lines were determined through the application of the high order moment $\upsilon\,T_{\rm mb}^3$ (to damp out the noise), i.e. $\upsilon_0 = \frac{\sum \upsilon\,T_{\rm mb}^3}{\sum T_{\rm mb}^3}$.

\begin{figure}
  \resizebox{\hsize}{!}{\includegraphics{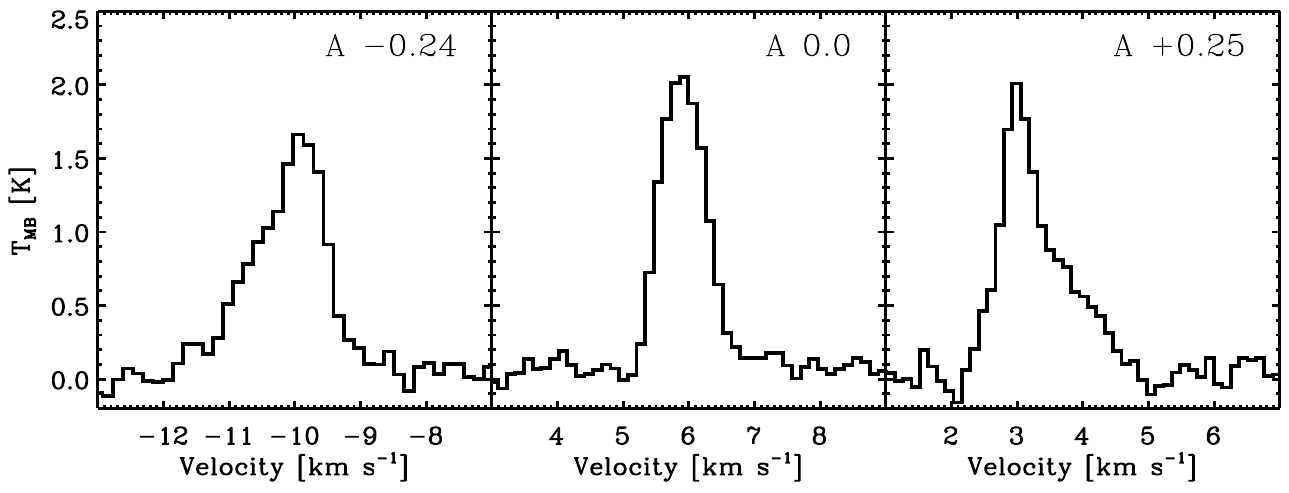}}
  \caption{Three types of observed CS\,(2-1) line profiles: blue-asymmetric (left), symmetric (centre) and red-asymmetric (right). Numerical values of the asymmetry parameter $A$ are given in the right upper corner of each panel.}
  \label{profiles}
\end{figure}

We then define an asymmetry parameter $A$ as the difference between the integrated intensity on the positive side and on the negative side of the line centre $\upsilon_0$, viz. $A = \frac{ \int_{\upsilon_0}^{\infty} T_{\rm mb}\,d\upsilon - \int_{-\infty}^{\upsilon_0} T_{\rm mb}\,d\upsilon} { \int_{-\infty}^{\infty} T_{\rm mb}\,d\upsilon}$ and $\mid \! \!A \! \!\mid \,\in \left [0,\,1 \right ]$. Examples of observed line asymmetries are shown in Fig.\,\ref{profiles}.

\begin{figure}
  \resizebox{\hsize}{!}{\includegraphics{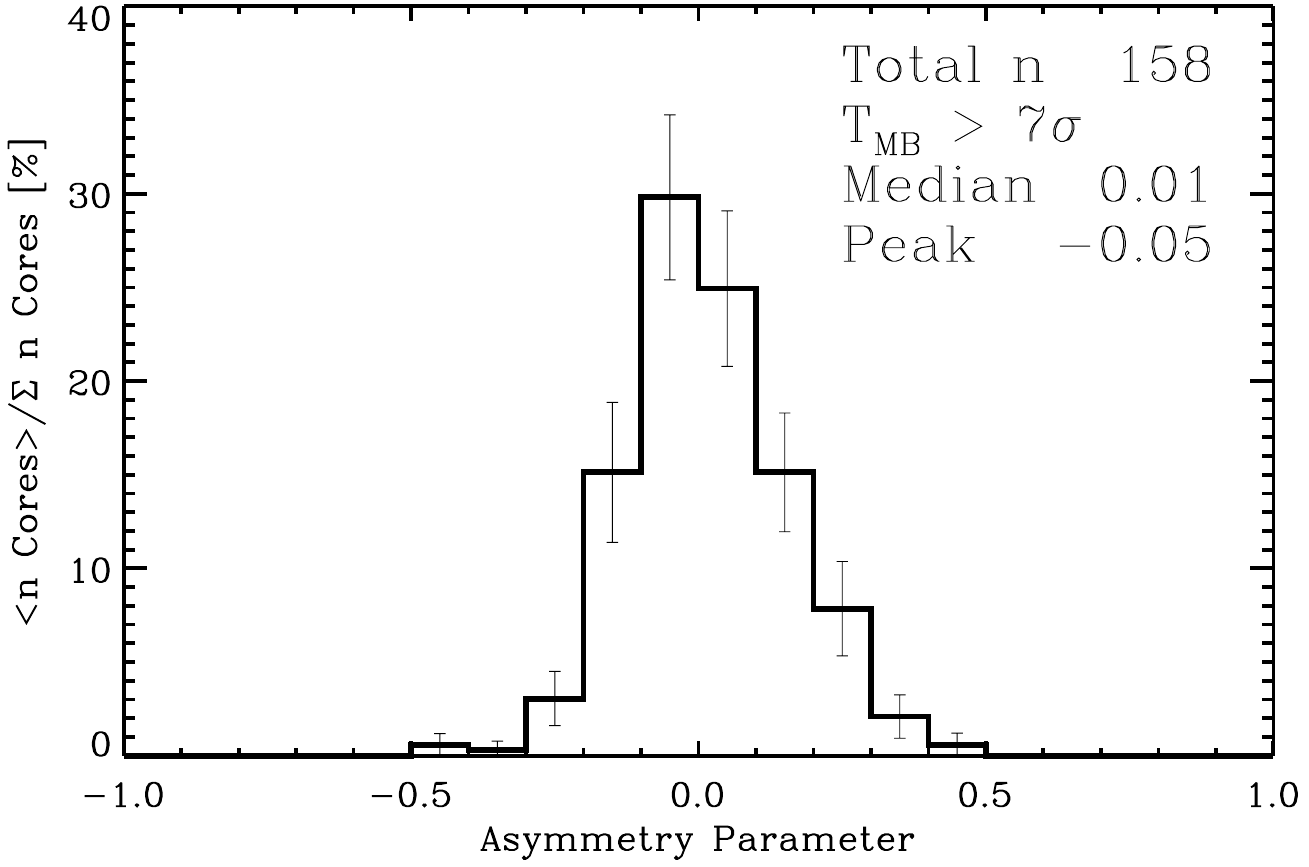}}
  \caption{The distribution of the asymmetry parameter $A$ observed toward 158 cores in the CS(2-1) line. For 17\%  (27 cores) $A < -0.1$ and for 24\% (38 cores) $A > 0.1 $.}
  \label{ass}
\end{figure}

The result for those 158 cores for which $T_{\rm mb,\,0} > 7\sigma$ is shown in Fig.\,\ref{ass} (obtained toward the position of maximum CS\,(2-1) emission). Only 3\% (5 cores) have a clear blue-asymmetry, with the asymmetry parameter $A < -0.2$, while a somewhat larger proportion, i.e. 11\% (17 cores), are clearly red-asymmetric with $A >0.2$. Evidently, the majority of cores reveals symmetric line profiles, with 41\% (65 cores) displaying $\mid \! \!A \! \!\mid \,\in \left [-0.1,\,0.1 \right ]$, perhaps indicating that outflows are not easily detectable in the lines of CS or have not yet developed in the cores.

\section{Conclusions}

Based on extensive mm-wave observations of a large sample of small dark cores in both celestial hemispheres in rotational lines of the CS molecule we conclude the following:

\begin{itemize}

\item[$\bullet$] The rate of CS detections is high. Specifically, out of 471 cores observed in CS\,(2-1), 315 were detected above $3\sigma$. For CS\,(3-2), these numbers are 431 and 141, respectively.
\item[$\bullet$] For 38\% of the detected cores does the optical centre coincide with the position of maximum CS emission.
\item[$\bullet$] Core temperatures are found to be low (\lapprox 10\,K).
\item[$\bullet$] On the scales of these observations, the majority of cores is optically thin in the CS emission, with the median column density $N({\rm CS}) = 7\times 10^{12}$\,\cmtwo.
\item[$\bullet$] The lines are generally symmetric and narrow, but wider than due to thermal broadening alone. The median line width is FWHM\,=\,0.7\,\kms. 
\end{itemize}

\begin{acknowledgements} We thank the referee, Dr. Jan Brand, for his valuable comments on the manuscript. Special thanks go also to S.\,Nasoudi-Shoar and F.\,Saunier for their contributions to the project. This work was supported  by the Swedish Science Research Council (NFR) and the Swedish National Space Board (SNSB). 
\end{acknowledgements}


\begin{table*}
  \caption{\label{tab_obs_cs} CS\,(2-1) and (3-2) observations of cores (for a description, see the end of the table). }
  \resizebox{\hsize}{!}{\rotatebox{0}{\includegraphics{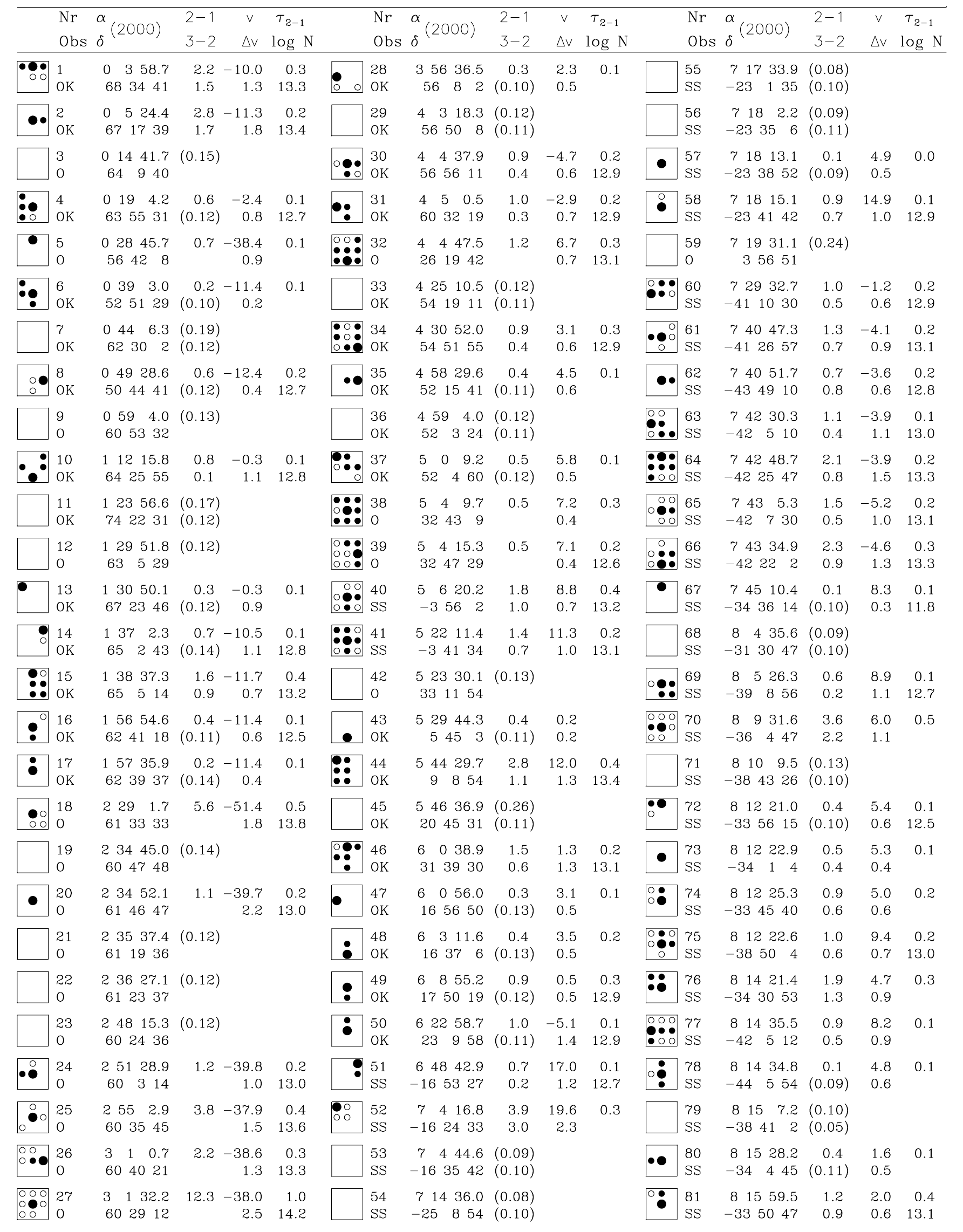}}}
\end{table*}
\addtocounter{table}{-1}
\begin{table*}
  \caption{Continued}
  \resizebox{\hsize}{!}{\rotatebox{0}{\includegraphics{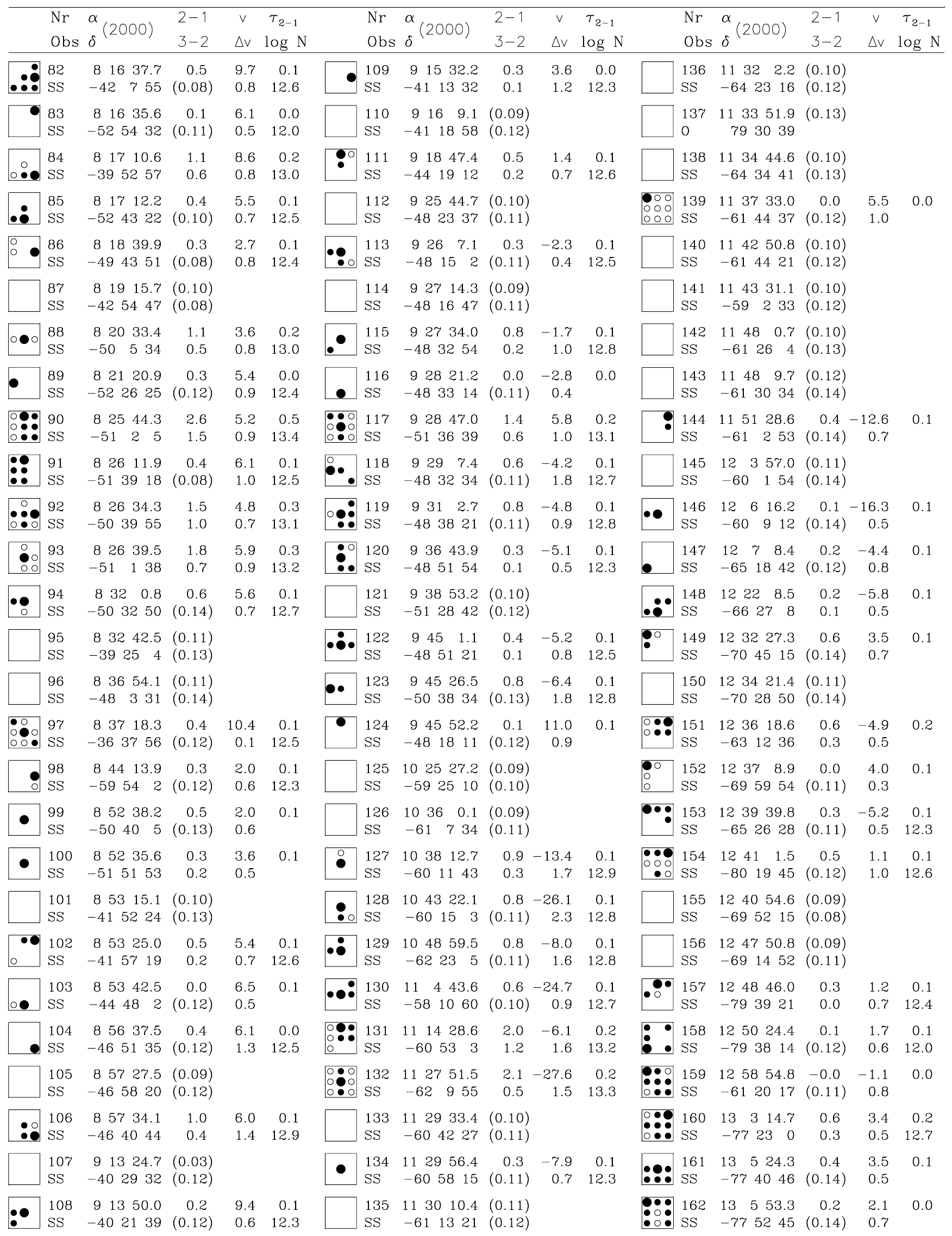}}}
\end{table*}
\addtocounter{table}{-1}
\begin{table*}
  \caption{Continued}
  \resizebox{\hsize}{!}{\rotatebox{0}{\includegraphics{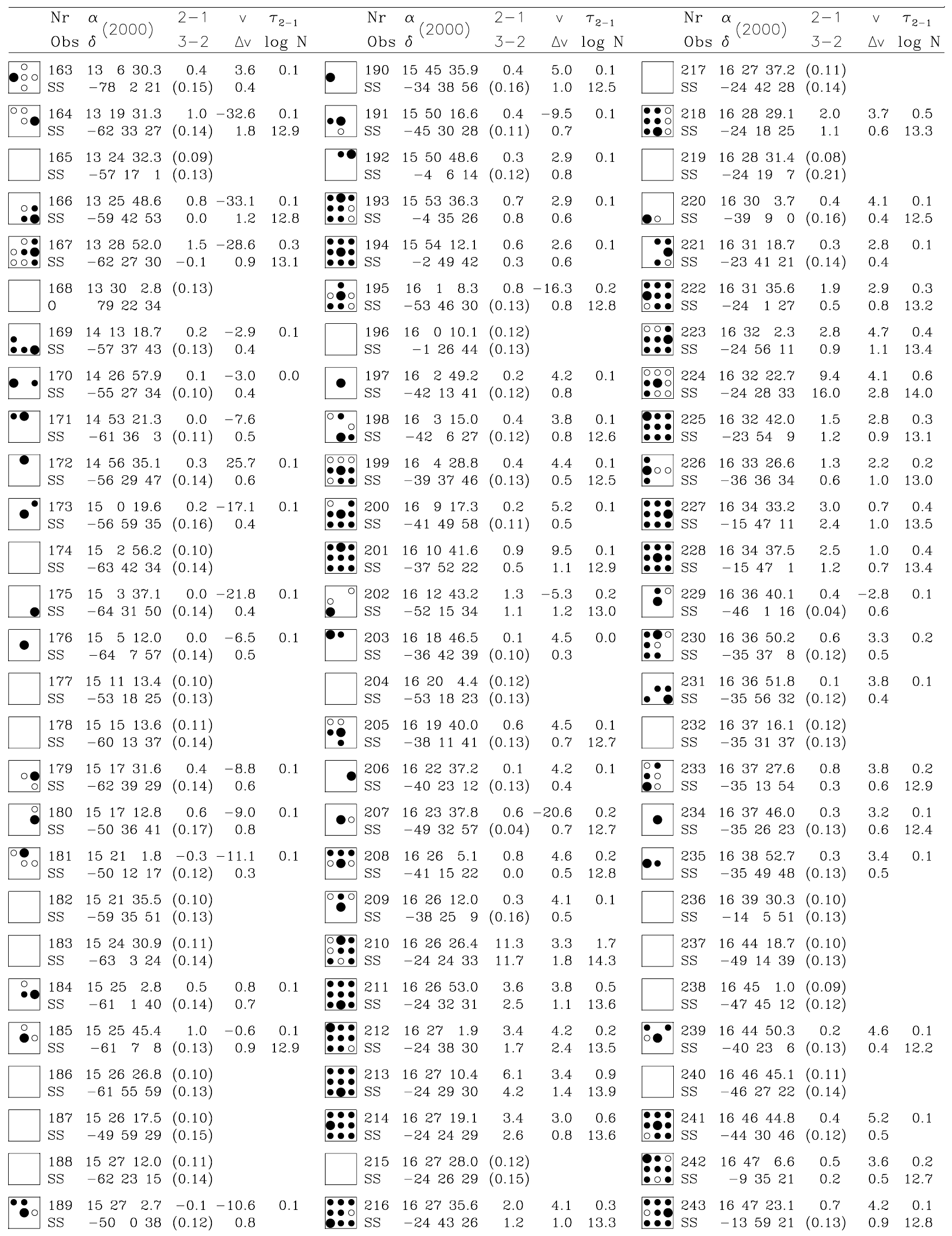}}}
\end{table*}
\addtocounter{table}{-1}
\begin{table*}
  \caption{Continued}
  \resizebox{\hsize}{!}{\rotatebox{0}{\includegraphics{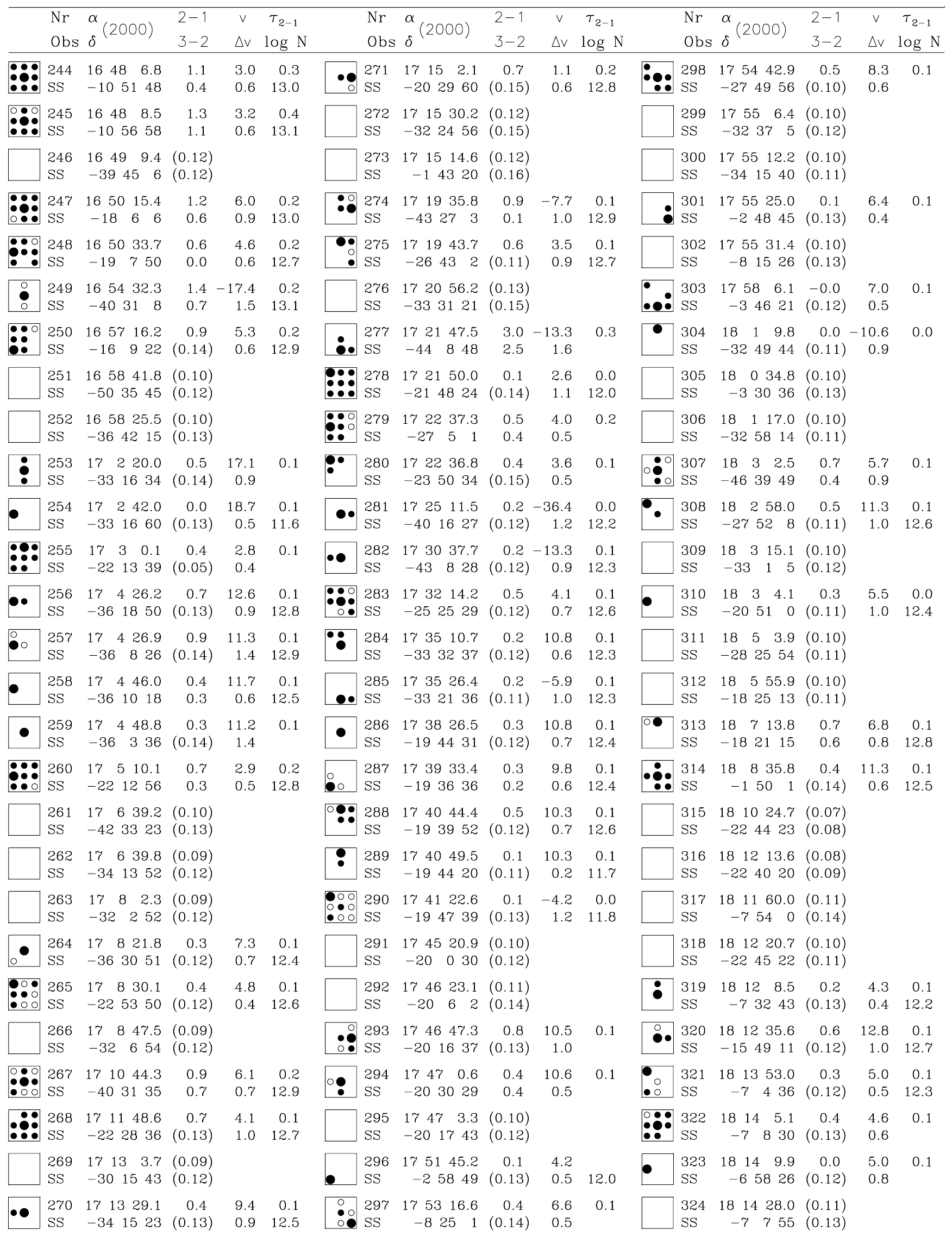}}}
\end{table*}
\addtocounter{table}{-1}
\begin{table*}
  \caption{Continued}
  \resizebox{\hsize}{!}{\rotatebox{0}{\includegraphics{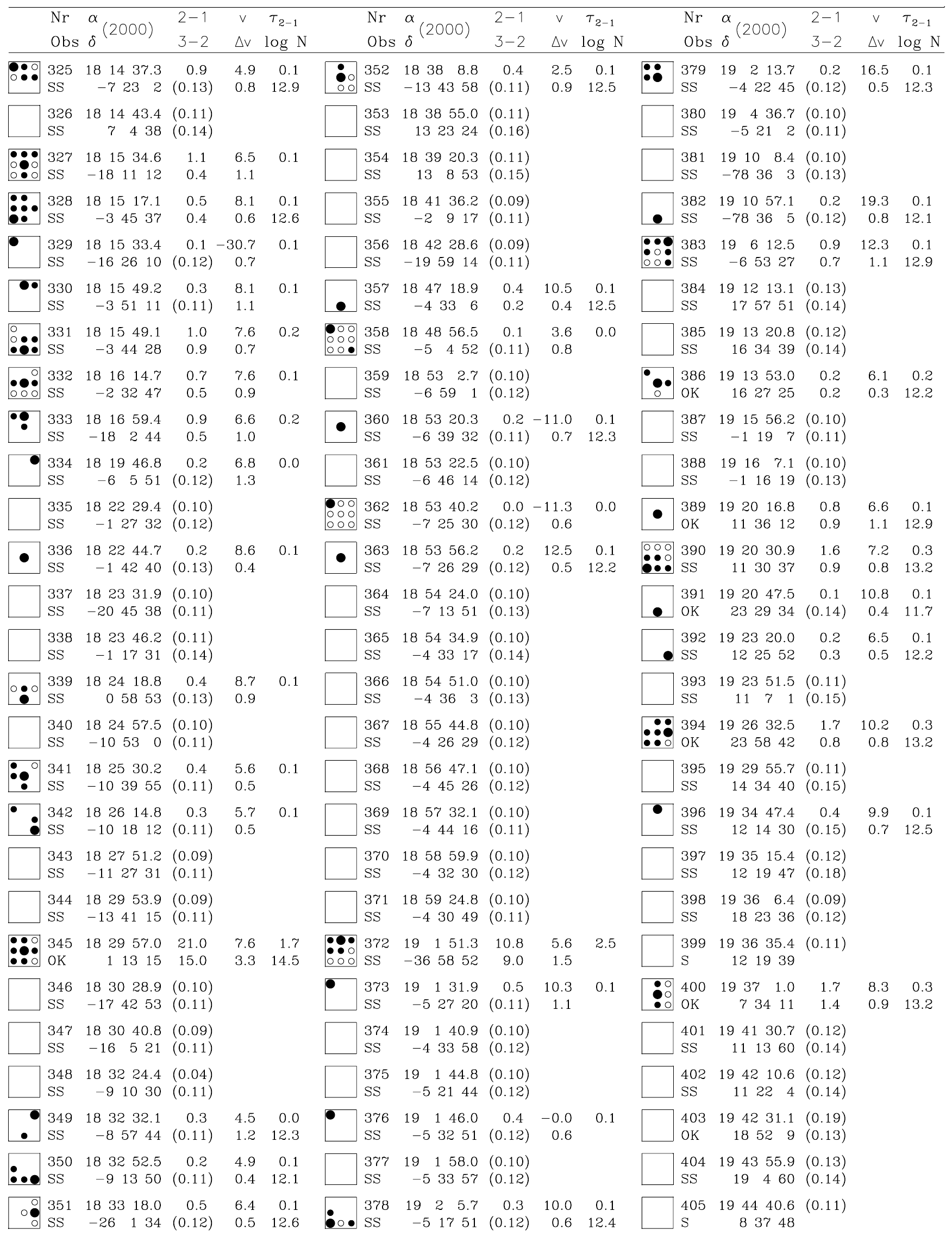}}}
\end{table*}
\addtocounter{table}{-1}
\begin{table*}
  \caption{Continued}
  \resizebox{\hsize}{!}{\rotatebox{0}{\includegraphics{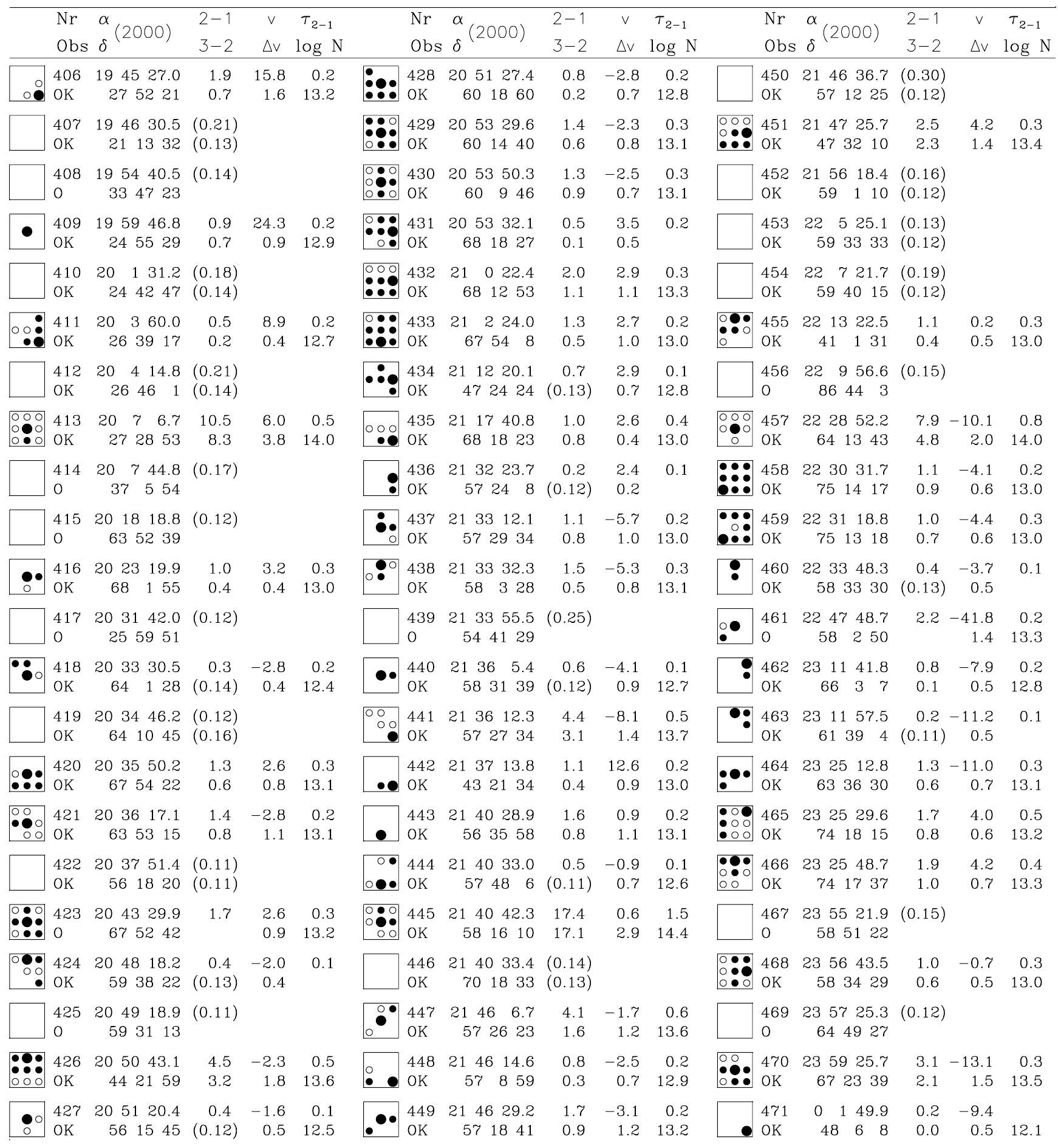}}}  
\\
Description of the table: \\
Column 1: A small 9 point, 1\amin\ spacings, map per core, where large filled dots refer to the position
of maximum CS\,(2-1) emission, small filled dots to emission above the 50\% level
of the maximum. Small open symbols denote positions with emission above the $3\sigma$ level.
Blanks are non-detections. \\
Column 2: In the upper row a running number is given, and in the lower the observatories are
identified: O = OSO, K = Kitt Peak, S = SEST. \\
Columns $3-5$: Right Ascension and Declination of epoch J2000 are given in the upper and 
lower row, respectively.\\
Column 6: The integrated line intensity $\int\!T_{\rm mb} dv$ in K\,\kms\ is given for CS\,(2-1)
in the upper row, and for CS\,(3-2) in the lower. For non-detections, the $T_{\rm mb}$-rms value (in K) is given in parentheses. \\
Column 7: The LSR-velocity (upper) and line width (lower) in \kms\ of the CS\,(2-1) line. \\
Column 8: The optical depth (upper) in the CS\,(2-1) line and the total CS column density in \cmtwo\
(lower).
\end{table*}


\begin{table*}
\caption{Core IDs, Coordinates, References and Observatories}
\begin{flushleft}
\begin{tabular}{rllcccc}
Nr  & RA (J2000) & Dec (J2000)       & Name               & Cat & Observatory & Observatory    \\
     & h  m  s   & \adeg\ \amin\ \asec   &                          &        & CS (2-1)       & CS (3-2)           \\  
\hline
  1 &  0  3 58.66  & 68 34 40.9 & Sug2 HII(NGC7822)  &  S91&    OSO 97 &  KITT 97 \\
  2 &  0  5 24.43  & 67 17 38.9 & Sug3 HII(NGC7822)  &  S91&    OSO 97 &  KITT 97 \\
  3 &  0 14 41.67  & 64  9 40.3 &         CB1 KHAV3  &  Cle&    OSO 97 &       no \\
  4 &  0 19  4.16  & 63 55 31.0 &         CB2 L1281  &  Cle&    OSO 97 &  KITT 97 \\
  5 &  0 28 45.75  & 56 42  7.9 &        CB3 LBN594  &  Cle&    OSO 97 &       no \\
  6 &  0 39  3.04  & 52 51 28.5 &            CB4 DC  &  Cle&    OSO  98&   KITT 97\\
  7 &  0 44  6.28  & 62 30  1.7 &         CB5 L1301  &  Cle&   OSO  98 &  KITT 97 \\
  8 &  0 49 28.62  & 50 44 41.1 &        CB6 LBN613  &  Cle&   OSO  98 &  KITT 97 \\
  9 &  0 59  4.03  & 60 53 31.6 &    Sug4 HII(S185)  &  S91&    OSO 97 &       no \\
 10 &  1 12 15.79  & 64 25 54.8 &              CB7   &  Cle&    OSO 97 & KITT 97  \\
 11 &  1 23 56.58  & 74 22 30.6 &              CB8   &  Cle&    OSO 97 &  KITT 97 \\
 12 &  1 29 51.81  & 63  5 28.8 &         CB9 L1325  &  Cle&    OSO 97 &       no \\
 13 &  1 30 50.09  & 67 23 46.3 &             CB10   &  Cle&    OSO 97 &  KITT 97 \\
 14 &  1 37  2.33  & 65  2 42.7 &             CB11   &  Cle&    OSO 97 &  KITT 97 \\
 15 &  1 38 37.33  & 65  5 13.9 &             CB12   &  Cle&   OSO  98 &  KITT 97 \\
 16 &  1 56 54.64  & 62 41 18.3 &        CB13 L1345  &  Cle&    OSO 97 &  KITT 97 \\
 17 &  1 57 35.90  & 62 39 36.8 &        CB14 L1346  &  Cle&    OSO 97 &  KITT 97 \\
 18 &  2 29  1.70  & 61 33 32.5 &    Sug5 HII(S190)  &  S91&    OSO 97 &       no \\
 19 &  2 34 44.99  & 60 47 48.3 &    Sug6 HII(S190)  &  S91&    OSO 97 &       no \\
 20 &  2 34 52.08  & 61 46 47.0 &    Sug7 HII(S190)  &  S91&    OSO 97 &       no \\
 21 &  2 35 37.37  & 61 19 36.0 &    Sug8 HII(S190)  &  S91&    OSO 97 &       no \\
 22 &  2 36 27.12  & 61 23 36.7 &    Sug9 HII(S190)  &  S91&    OSO 97 &       no \\
 23 &  2 48 15.33  & 60 24 35.5 &   Sug10 HII(S199)  &  S91&    OSO 97 &       no \\
 24 &  2 51 28.93  & 60  3 14.1 &   Sug11 HII(S199)  &  S91&    OSO 97 &       no \\
 25 &  2 55  2.90  & 60 35 44.6 &   Sug12 HII(S199)  &  S91&    OSO 97 &       no \\
 26 &  3  1  0.67  & 60 40 20.5 &   Sug13 HII(S199)  &  S91&    OSO 97 &       no \\
 27 &  3  1 32.24  & 60 29 11.9 &   Sug14 HII(S199)  &  S91&    OSO 97 &       no \\
 28 &  3 56 36.47  & 56  8  2.2 &     CB15 B6,L1387  &  Cle&    OSO 97 &  KITT 97 \\
 29 &  4  3 18.27  & 56 50  8.1 &        CB16 L1388  &  Cle&    OSO 97 &  KITT 97 \\
 30 &  4  4 37.90  & 56 56 11.1 &        CB17 L1389  &  Cle&    OSO 97 &  KITT 97 \\
 31 &  4  5  0.55  & 60 32 19.1 &             CB18   &  Cle&    OSO 97 &  KITT 97 \\
 32 &  4  4 47.46  & 26 19 41.6 &             L1489  &  BM &    OSO 97 &       no \\
 33 &  4 25 10.46  & 54 19 10.8 &              L14G  &  BM &    OSO 97 &  KITT 97 \\
 34 &  4 30 52.03  & 54 51 55.1 &              L14K  &  BM &    OSO 97 &  KITT 97 \\
 35 &  4 58 29.59  & 52 15 41.0 &             CB24   &  Cle&    OSO 97 &  KITT 97 \\
 36 &  4 59  4.05  & 52  3 23.6 &        CB25 L1437  &  Cle&    OSO 97 &  KITT 97 \\
 37 &  5  0  9.24  & 52  4 60.0 &        CB26 L1439  &  Cle&    OSO 97 &  KITT 97 \\
 38 &  5  4  9.71  & 32 43  8.6 &             L1512  &  BM &    OSO 97 &       no \\
 39 &  5  4 15.33  & 32 47 29.2 &       CB27 SCHO93  &  Cle&    OSO 97 &       no \\
 40 &  5  6 20.19  & -3 56  2.2 &   CB28     LBN923  &  Cle&   SEST 96 &  SEST 96 \\
 41 &  5 22 11.41  & -3 41 34.1 &              CB29  &  Cle&   SEST 96 &  SEST 96 \\
 42 &  5 23 30.06  & 33 11 54.0 &   Sug15 HII(S236)  &  S91&    OSO 97 &       no \\
 43 &  5 29 44.33  &  5 45  2.7 &             CB30   &  Cle&   OSO  98 &  KITT 97 \\
 44 &  5 44 29.70  &  9  8 53.7 &        B35A,Sug18  &  BM &   OSO  98 &  KITT 97 \\
 45 &  5 46 36.94  & 20 45 31.0 &             CB33   &  Cle&   OSO  98 &  KITT 97 \\
 46 &  6  0 38.88  & 31 39 30.3 &        CB37 L1555  &  Cle&    OSO 97 &  KITT 97 \\
 47 &  6  0 56.04  & 16 56 50.3 &             CB38   &  Cle&   OSO  98 &  KITT 97 \\
 48 &  6  3 11.63  & 16 37  6.4 &             CB43   &  Cle&   OSO  98 &  KITT 97 \\
 49 &  6  8 55.18  & 17 50 19.4 &      CB45 L1578(?)  &  Cle&   OSO  98 &  KITT 97 \\
 50 &  6 22 58.66  & 23  9 58.3 &   Sug23 HII(S249)  &  S91&    OSO 97 &  KITT 97 \\
 51 &  6 48 42.87  &-16 53 26.6 &              CB52  &  Cle&   SEST 96 &  SEST 96 \\
 52 &  7  4 16.80  &-16 24 32.6 &  CB54     LBN1042  &  Cle&   SEST 96 &  SEST 96 \\
 53 &  7  4 44.59  &-16 35 41.6 &              CB55  &  Cle&   SEST 96 &  SEST 96 \\
 54 &  7 14 35.99  &-25  8 54.1 &              CB56  &  Cle&   SEST 96 &  SEST 96 \\
 55 &  7 17 33.92  &-23  1 35.3 &              CB57  &  Cle&   SEST 96 &  SEST 96 \\
 56 &  7 18  2.21  &-23 35  6.3 &            CB58-1  &  Cle&   SEST 96 &  SEST 96 \\
 57 &  7 18 13.14  &-23 38 52.0 &              CB58  &  Cle&   SEST 96 &  SEST 96 \\
 58 &  7 18 15.08  &-23 41 42.2 &            CB58-2  &  Cle&   SEST 96 &  SEST 96 \\
\end{tabular}
\end{flushleft}
\end{table*}
\clearpage
\newpage

\begin{table*}
\begin{flushleft}
\begin{tabular}{rllcccc}
Nr  & RA (J2000) & Dec (J2000)       & Name               & Cat & Observatory & Observatory    \\
     & h  m  s   & \adeg\ \amin\ \asec   &                          &        & CS (2-1)       & CS (3-2)           \\  
\hline
 59 &  7 19 31.08  &  3 56 50.7 &             CB59   &  Cle&    OSO 97 &       no \\
 60 &  7 29 32.68  &-41 10 30.3 &     DC 253.8-10.9  &  Har&   SEST 96 &  SEST 96 \\
 61 &  7 40 47.32  &-41 26 57.3 &      DC 255.1-9.2  &  Har&   SEST 96 &  SEST 96 \\
 62 &  7 40 51.72  &-43 49  9.7 &     DC 257.2-10.3  &  Har&   SEST 96 &  SEST 96 \\
 63 &  7 42 30.31  &-42  5 10.1 &      DC 255.8-9.2  &  Har&   SEST 96 &  SEST 96 \\
 64 &  7 42 48.69  &-42 25 47.3 &      DC 256.1-9.3  &  Har&   SEST 96 &  SEST 96 \\
 65 &  7 43  5.30  &-42  7 30.4 &      DC 255.9-9.1  &  Har&   SEST 96 &  SEST 96 \\
 66 &  7 43 34.90  &-42 22  2.4 &      DC 256.1-9.1  &  Har&   SEST 96 &  SEST 96 \\
 67 &  7 45 10.38  &-34 36 14.2 &      DC 249.4-5.1  &  Har&   SEST 96 &  SEST 96 \\
 68 &  8  4 35.64  &-31 30 46.6 &    CB60     L1670  &  Cle&   SEST 96 &  SEST 96 \\
 69 &  8  5 26.26  &-39  8 56.1 &      DC 255.4-3.9  &  Har&   SEST 96 &  SEST 96 \\
 70 &  8  9 31.58  &-36  4 47.2 &      DC 253.3-1.6  &  Har&   SEST 96 &  SEST 96 \\
 71 &  8 10  9.54  &-38 43 25.6 &      DC 255.6-2.9  &  Har&   SEST 96 &  SEST 96 \\
 72 &  8 12 21.03  &-33 56 15.5 &      DC 251.8+0.0  &  Har&   SEST 96 &  SEST 96 \\
 73 &  8 12 22.92  &-34  1  3.6 &      DC 251.9+0.0  &  Har&   SEST 96 &  SEST 96 \\
 74 &  8 12 25.29  &-33 45 39.7 &      DC 251.7+0.2  &  Har&   SEST 96 &  SEST 96 \\
 75 &  8 12 22.64  &-38 50  3.8 &      DC 255.9-2.6  &  Har&   SEST 96 &  SEST 96 \\
 76 &  8 14 21.42  &-34 30 52.8 &      DC 252.5+0.1  &  Har&   SEST 96 &  SEST 96 \\
 77 &  8 14 35.47  &-42  5 12.1 &      DC 258.9-4.1  &  Har&   SEST 96 &  SEST 96 \\
 78 &  8 14 34.81  &-44  5 54.1 &      DC 260.5-5.2  &  Har&   SEST 96 &  SEST 96 \\
 79 &  8 15  7.23  &-38 41  1.8 &      DC 256.1-2.1  &  Har&   SEST 96 & SEST 96  \\
 80 &  8 15 28.17  &-34  4 44.9 &      DC 252.3+0.5  &  Har&   SEST 96 &  SEST 96 \\
 81 &  8 15 59.55  &-33 50 46.8 &      DC 252.2+0.7  &  Har&   SEST 96 &  SEST 96 \\
 82 &  8 16 37.69  &-42  7 55.5 &      DC 259.1-3.8  &  Har&   SEST 96 &  SEST 96 \\
 83 &  8 16 35.63  &-52 54 32.0 &      DC 268.2-9.7  &  Har&   SEST 96 &  SEST 96 \\
 84 &  8 17 10.58  &-39 52 57.3 &      DC 257.3-2.5  &  Har&   SEST 96 &  SEST 96 \\
 85 &  8 17 12.25  &-52 43 22.2 &      DC 268.1-9.5  &  Har&   SEST 96 &  SEST 96 \\
 86 &  8 18 39.87  &-49 43 51.3 &      DC 265.7-7.7  &  Har&   SEST 96 &  SEST 96 \\
 87 &  8 19 15.71  &-42 54 47.0 &      DC 260.0-3.8  &  Har&   SEST 96 &  SEST 96 \\
 88 &  8 20 33.41  &-50  5 34.1 &      DC 266.1-7.7  &  Har&   SEST 96 &  SEST 96 \\
 89 &  8 21 20.88  &-52 26 25.1 &      DC 268.2-8.9  &  Har&   SEST 96 &  SEST 96 \\
 90 &  8 25 44.28  &-51  2  4.5 &      DC 267.4-7.5  &  Har&   SEST 96 &  SEST 96 \\
 91 &  8 26 11.89  &-51 39 18.2 &      DC 267.9-7.8  &  Har&   SEST 96 &  SEST 96 \\
 92 &  8 26 34.32  &-50 39 55.4 &      DC 267.2-7.2  &  Har&   SEST 96 &  SEST 96 \\
 93 &  8 26 39.49  &-51  1 37.8 &      DC 267.5-7.4  &  Har&   SEST 96 &  SEST 96 \\
 94 &  8 32  0.75  &-50 32 50.4 &      DC 267.6-6.4  &  Har&   SEST 96 &  SEST 96 \\
 95 &  8 32 42.52  &-39 25  4.2 &      DC 258.6+0.3  &  Har&   SEST 96 &  SEST 96 \\
 96 &  8 36 54.11  &-48  3 30.9 &      DC 266.0-4.3  &  Har&   SEST 96 &  SEST 96 \\
 97 &  8 37 18.25  &-36 37 55.8 &      DC 256.9+2.6  &  Har&   SEST 96 &  SEST 96 \\
 98 &  8 44 13.92  &-59 54  2.3 &     DC 276.2-10.6  &  Har&   SEST 96 &  SEST 96 \\
 99 &  8 52 38.23  &-50 40  4.9 &      DC 269.7-3.9  &  Har&   SEST 96 &  SEST 96 \\
100 &  8 52 35.63  &-51 51 52.8 &      DC 270.6-4.7  &  Har&   SEST 96 &  SEST 96 \\
101 &  8 53 15.10  &-41 52 24.4 &      DC 263.0+1.8  &  Har&   SEST 96 &  SEST 96 \\
102 &  8 53 25.00  &-41 57 19.0 &      DC 263.1+1.8  &  Har&   SEST 96 &  SEST 96 \\
103 &  8 53 42.50  &-44 48  2.0 &      DC 265.3-0.0  &  Har&   SEST 96 &  SEST 96 \\
104 &  8 56 37.53  &-46 51 35.3 &      DC 267.2-0.0  &  Har&   SEST 96 &  SEST 96 \\
105 &  8 57 27.51  &-46 58 20.0 &      DC 267.4-0.9  &  Har&   SEST 96 &  SEST 96 \\
106 &  8 57 34.06  &-46 40 44.3 &      DC 267.1-0.7  &  Har&   SEST 96 &  SEST 96 \\
107 &  9 13 24.75  &-40 29 32.0 &      DC 264.5+5.6  &  Har&   SEST 96 &  SEST 96 \\
108 &  9 13 50.00  &-40 21 39.2 &      DC 264.4+5.7  &  Har&   SEST 96 &  SEST 96 \\
109 &  9 15 32.15  &-41 13 32.1 &      DC 265.3+5.3  &  Har&   SEST 96 &  SEST 96 \\
110 &  9 16  9.15  &-41 18 57.9 &      DC 265.4+5.4  &  Har&   SEST 96 &  SEST 96 \\
111 &  9 18 47.45  &-44 19 11.5 &      DC 267.9+3.6  &  Har&   SEST 96 &  SEST 96 \\
112 &  9 25 44.75  &-48 23 37.1 &      DC 271.6+1.6  &  Har&   SEST 96 &  SEST 96 \\
113 &  9 26  7.08  &-48 15  2.1 &      DC 271.6+1.7  &  Har&   SEST 96 &  SEST 96 \\
114 &  9 27 14.31  &-48 16 47.1 &      DC 271.7+1.8  &  Har&   SEST 96 &  SEST 96 \\
115 &  9 27 33.96  &-48 32 54.0 &      DC 272.0+1.7  &  Har&   SEST 96 &  SEST 96 \\
116 &  9 28 21.15  &-48 33 14.1 &      DC 272.1+1.8  &  Har&   SEST 96 &  SEST 96 \\
\end{tabular}
\end{flushleft}
\end{table*}
\clearpage
\newpage

\begin{table*}
\begin{flushleft}
\begin{tabular}{rllcccc}
Nr  & RA (J2000) & Dec (J2000)       & Name               & Cat & Observatory & Observatory    \\
     & h  m  s   & \adeg\ \amin\ \asec   &                          &        & CS (2-1)       & CS (3-2)           \\  
\hline
117 &  9 28 46.96  &-51 36 39.4 &      DC 274.2-0.4  &  Har&   SEST 96 &  SEST 96 \\
118 &  9 29  7.37  &-48 32 34.2 &      DC 272.2+1.9  &  Har&   SEST 96 &  SEST 96 \\
119 &  9 31  2.71  &-48 38 21.3 &      DC 272.5+2.0  &  Har&   SEST 96 &  SEST 96 \\
120 &  9 36 43.86  &-48 51 54.2 &      DC 273.3+2.5  &  Har&   SEST 96 &  SEST 96 \\
121 &  9 38 53.16  &-51 28 41.8 &      DC 275.3+0.8  &  Har&   SEST 96 &  SEST 96 \\
122 &  9 45  1.13  &-48 51 20.9 &      DC 274.3+3.4  &  Har&   SEST 96 &  SEST 96 \\
123 &  9 45 26.48  &-50 38 34.0 &      DC 275.5+2.1  &  Har&   SEST 96 &  SEST 96 \\
124 &  9 45 52.18  &-48 18 11.0 &      DC 274.1+3.9  &  Har&   SEST 96 &  SEST 96 \\
125 & 10 25 27.19  &-59 25 10.4 &      DC 285.3-1.6  &  Har&   SEST 96 &  SEST 96 \\
126 & 10 36  0.06  &-61  7 34.0 &      DC 287.3-2.4  &  Har&   SEST 96 &  SEST 96 \\
127 & 10 38 12.72  &-60 11 43.4 &      DC 287.1-1.5  &  Har&   SEST 96 &  SEST 96 \\
128 & 10 43 22.07  &-60 15  3.1 &      DC 287.7-1.2  &  Har&   SEST 96 &  SEST 96 \\
129 & 10 48 59.47  &-62 23  5.0 &      DC 289.3-2.8  &  Har&   SEST 96 &  SEST 96 \\
130 & 11  4 43.62  &-58 10 59.9 &      DC 289.2+1.8  &  Har&   SEST 96 &  SEST 96 \\
131 & 11 14 28.60  &-60 53  3.4 &      DC 291.4-0.2  &  Har&   SEST 96 &  SEST 96 \\
132 & 11 27 51.49  &-62  9 55.5 &      DC 293.3-0.9  &  Har&   SEST 96 &  SEST 96 \\
133 & 11 29 33.42  &-60 42 26.5 &      DC 293.1+0.6  &  Har&   SEST 96 &  SEST 96 \\
134 & 11 29 56.44  &-60 58 14.7 &      DC 293.2+0.4  &  Har&   SEST 96 &  SEST 96 \\
135 & 11 30 10.39  &-61 13 20.9 &      DC 293.3+0.1  &  Har&   SEST 96 &  SEST 96 \\
136 & 11 32  2.15  &-64 23 16.0 &      DC 294.5-2.8  &  Har&   SEST 96 &  SEST 96 \\
137 & 11 33 51.92  & 79 30 39.3 &             CB61   &  Cle&    OSO 97 &       no \\
138 & 11 34 44.63  &-64 34 41.4 &      DC 294.8-2.9  &  Har&   SEST 96 &  SEST 96 \\
139 & 11 37 32.97  &-61 44 36.7 &      DC 294.3-0.1  &  Har&  SEST 96  &  SEST 96 \\
140 & 11 42 50.80  &-61 44 20.8 &      DC 294.9+0.1  &  Har&   SEST 96 &  SEST 96 \\
141 & 11 43 31.14  &-59  2 33.0 &      DC 294.3+2.7  &  Har&   SEST 96 &  SEST 96 \\
142 & 11 48  0.66  &-61 26  4.4 &      DC 295.4+0.5  &  Har&   SEST 96 &  SEST 96 \\
143 & 11 48  9.72  &-61 30 34.4 &      DC 295.5+0.4  &  Har&   SEST 96 &  SEST 96 \\
144 & 11 51 28.58  &-61  2 53.1 &      DC 295.7+0.0  &  Har&   SEST 96 &  SEST 96 \\
145 & 12  3 57.02  &-60  1 54.0 &      DC 297.0+2.3  &  Har&   SEST 96 &  SEST 96 \\
146 & 12  6 16.20  &-60  9 11.8 &      DC 297.3+2.2  &  Har&   SEST 96 &  SEST 96 \\
147 & 12  7  8.36  &-65 18 41.7 &      DC 298.3-2.8  &  Har&   SEST 96 &  SEST 96 \\
148 & 12 22  8.45  &-66 27  7.9 &      DC 300.0-3.7  &  Har&   SEST 96 &  SEST 96 \\
149 & 12 32 27.26  &-70 45 14.9 &      DC 301.4-7.9  &  Har&   SEST 96 &  SEST 96 \\
150 & 12 34 21.40  &-70 28 49.7 &      DC 301.5-7.7  &  Har&   SEST 96 &  SEST 96 \\
151 & 12 36 18.61  &-63 12 36.4 &      DC 301.2-0.4  &  Har&   SEST 96 &  SEST 96 \\
152 & 12 37  8.90  &-69 59 53.9 &      DC 301.7-7.2  &  Har&   SEST 96 &  SEST 96 \\
153 & 12 39 39.77  &-65 26 28.2 &      DC 301.7-2.6  &  Har&   SEST 96 &  SEST 96 \\
154 & 12 41  1.45  &-80 19 45.4 &     DC 302.5-17.5  &  Har&   SEST 96 &  SEST 96 \\
155 & 12 40 54.61  &-69 52 15.3 &      DC 302.0-7.0  &  Har&   SEST 96 &  SEST 96 \\
156 & 12 47 50.79  &-69 14 51.7 &      DC 302.6-6.4  &  Har&   SEST 96 &  SEST 96 \\
157 & 12 48 46.03  &-79 39 21.1 &     DC 302.8-16.8  &  Har&   SEST 96 & SEST 96  \\
158 & 12 50 24.39  &-79 38 13.7 &     DC 302.9-16.8  &  Har&   SEST 96 &  SEST 96 \\
159 & 12 58 54.80  &-61 20 16.8 &      DC 303.8+1.5  &  Har&  SEST 96  &  SEST 96 \\
160 & 13  3 14.73  &-77 23  0.4 &     DC 303.6-14.5  &  Har&   SEST 96 &  SEST 96 \\
161 & 13  5 24.34  &-77 40 45.9 &     DC 303.7-14.8  &  Har&   SEST 96 &  SEST 96 \\
162 & 13  5 53.33  &-77 52 45.3 &     DC 303.7-15.0  &  Har&  SEST 96  &  SEST 96 \\
163 & 13  6 30.27  &-78  2 20.6 &     DC 303.7-15.2  &  Har&   SEST 96 &  SEST 96 \\
164 & 13 19 31.33  &-62 33 26.7 &      DC 306.2+0.1  &  Har&   SEST 96 &  SEST 96 \\
165 & 13 24 32.33  &-57 17  1.1 &      DC 307.4+5.3  &  Har&   SEST 97 &  SEST 97 \\
166 & 13 25 48.55  &-59 42 53.2 &      DC 307.3+2.9  &  Har&   SEST 98 &  SEST 98 \\
167 & 13 28 51.99  &-62 27 30.4 &      DC 307.2+0.1  &  Har&   SEST 96 &  SEST 96 \\
168 & 13 30  2.82  & 79 22 34.1 &             CB62   &  Cle&    OSO 97 &       no \\
169 & 14 13 18.65  &-57 37 43.4 &      DC 313.7+3.5  &  Har&   SEST 97 &  SEST 97 \\
170 & 14 26 57.91  &-55 27 33.5 &      DC 316.3+4.9  &  Har&   SEST 97 &  SEST 97 \\
171 & 14 53 21.35  &-61 36  2.7 &      DC 316.9-2.1  &  Har&   SEST 96 &  SEST 96 \\
172 & 14 56 35.12  &-56 29 46.7 &      DC 319.6+2.3  &  Har&   SEST 97 &  SEST 97 \\
173 & 15  0 19.63  &-56 59 35.3 &      DC 319.8+1.6  &  Har&   SEST 97 &  SEST 97 \\
174 & 15  2 56.24  &-63 42 33.9 &      DC 316.9-4.5  &  Har&   SEST 96 &  SEST 96 \\
\end{tabular}
\end{flushleft}
\end{table*}
\clearpage
\newpage

\begin{table*}
\begin{flushleft}
\begin{tabular}{rllcccc}
Nr  & RA (J2000) & Dec (J2000)       & Name               & Cat & Observatory & Observatory    \\
     & h  m  s   & \adeg\ \amin\ \asec   &                          &        & CS (2-1)       & CS (3-2)           \\  
\hline
175 & 15  3 37.07  &-64 31 49.9 &      DC 316.6-5.2  &  Har&   SEST 96 &  SEST 96 \\
176 & 15  5 11.99  &-64  7 56.9 &      DC 316.9-4.9  &  Har&   SEST 96 &  SEST 96 \\
177 & 15 11 13.39  &-53 18 24.9 &      DC 323.0+4.0  &  Har&   SEST 97 &  SEST 97 \\
178 & 15 15 13.55  &-60 13 36.5 &      DC 319.9-2.2  &  Har&   SEST 96 &  SEST 96 \\
179 & 15 17 31.60  &-62 39 29.2 &      DC 318.8-4.4  &  Har&   SEST 96 &  SEST 96 \\
180 & 15 17 12.76  &-50 36 41.3 &      DC 325.2+5.8  &  Har&   SEST 97 &  SEST 97 \\
181 & 15 21  1.75  &-50 12 16.6 &      DC 325.9+5.9  &  Har&   SEST 97 &  SEST 97 \\
182 & 15 21 35.46  &-59 35 51.5 &      DC 320.9-2.1  &  Har&   SEST 96 &  SEST 96 \\
183 & 15 24 30.85  &-63  3 24.0 &      DC 319.3-5.2  &  Har&   SEST 96 &  SEST 96 \\
184 & 15 25  2.78  &-61  1 40.0 &      DC 320.5-3.5  &  Har&   SEST 96 &  SEST 96 \\
185 & 15 25 45.36  &-61  7  7.6 &      DC 320.5-3.6  &  Har&   SEST 96 &  SEST 96 \\
186 & 15 26 26.81  &-61 55 59.4 &      DC 320.1-4.3  &  Har&   SEST 96 &  SEST 96 \\
187 & 15 26 17.45  &-49 59 28.9 &      DC 326.8+5.6  &  Har&   SEST 98 &  SEST 98 \\
188 & 15 27 11.95  &-62 23 14.9 &      DC 319.9-4.8  &  Har&   SEST 96 &  SEST 96 \\
189 & 15 27  2.66  &-50  0 38.3 &      DC 326.9+5.5  &  Har&   SEST 97 &  SEST 97 \\
190 & 15 45 35.85  &-34 38 56.3 &     DC 339.0+15.8  &  Har&   SEST 98 &  SEST 98 \\
191 & 15 50 16.65  &-45 30 27.8 &      DC 332.7+6.8  &  Har&   SEST 97 &  SEST 97 \\
192 & 15 50 48.65  & -4  6 14.2 &     CB63    LBN11  &  Cle&   SEST 97 &  SEST 97 \\
193 & 15 53 36.26  & -4 35 25.9 &             L134A  &  BM &   SEST 97 &  SEST 97 \\
194 & 15 54 12.12  & -2 49 41.7 &             L183B  &  BM &   SEST 97 &  SEST 97 \\
195 & 16  1  8.30  &-53 46 30.0 &      DC 328.7-0.7  &  Har&   SEST 97 &  SEST 97 \\
196 & 16  0 10.07  & -1 26 44.3 &     CB64    LBN37  &  Cle&   SEST 97 &  SEST 97 \\
197 & 16  2 49.21  &-42 13 40.8 &      DC 336.6+7.8  &  Har&   SEST 97 &  SEST 97 \\
198 & 16  3 15.05  &-42  6 27.2 &      DC 336.7+7.8  &  Har&   SEST 97 &  SEST 97 \\
199 & 16  4 28.81  &-39 37 46.4 &      DC 338.6+9.5  &  Har&   SEST 97 &  SEST 97 \\
200 & 16  9 17.31  &-41 49 58.1 &      DC 337.7+7.3  &  Har&   SEST 97 &  SEST 97 \\
201 & 16 10 41.56  &-37 52 22.4 &      DC 340.7+0.0  &  Har&   SEST 97 &  SEST 97 \\
202 & 16 12 43.20  &-52 15 33.6 &      DC 331.0-0.7  &  Har&   SEST 97 &  SEST 97 \\
203 & 16 18 46.46  &-36 42 38.9 &      DC 342.7+9.7  &  Har&   SEST 97 &  SEST 97 \\
204 & 16 20  4.44  &-53 18 22.9 &      DC 331.1-2.3  &  Har&   SEST 97 &  SEST 97 \\
205 & 16 19 40.02  &-38 11 41.4 &      DC 341.7+8.5  &  Har&   SEST 97 &  SEST 97 \\
206 & 16 22 37.18  &-40 23 11.9 &      DC 340.6+6.6  &  Har&   SEST 97 &  SEST 97 \\
207 & 16 23 37.81  &-49 32 56.5 &      DC 334.2+0.0  &  Har&   SEST 97 & SEST 97  \\
208 & 16 26  5.15  &-41 15 22.1 &      DC 340.4+5.5  &  Har&   SEST 98 &  SEST 98 \\
209 & 16 26 12.04  &-38 25  9.5 &      DC 342.5+7.4  &  Har&   SEST 97 &  SEST 97 \\
210 & 16 26 26.36  &-24 24 32.8 &  L1688A(DCO+ 2-1)  &  Lor&   SEST 97 &  SEST 97 \\
211 & 16 26 52.95  &-24 32 31.1 &  L1688C(DCO+ 2-1)  &  Lor&   SEST 97 &  SEST 97 \\
212 & 16 27  1.89  &-24 38 30.5 &  L1688E(DCO+ 2-1)  &  Lor&   SEST 97 &  SEST 97 \\
213 & 16 27 10.41  &-24 29 29.9 & L1688B1(DCO+ 2-1)  &  Lor&   SEST 97 &  SEST 97 \\
214 & 16 27 19.11  &-24 24 29.3 & L1688B3(DCO+ 2-1)  &  Lor&   SEST 97 &  SEST 97 \\
215 & 16 27 27.96  &-24 26 28.7 & L1688B2(DCO+ 2-1)  &  Lor&   SEST 98 &  SEST 98 \\
216 & 16 27 35.63  &-24 43 26.2 &            L1681B  &  BM &   SEST 97 &  SEST 97 \\
217 & 16 27 37.21  &-24 42 28.1 &  L1688F(DCO+ 2-1)  &  Lor&   SEST 98 &  SEST 98 \\
218 & 16 28 29.14  &-24 18 24.6 &  L1688D(DCO+ 2-1)  &  Lor&   SEST 97 &  SEST 97 \\
219 & 16 28 31.35  &-24 19  7.5 &            L1696A  &  BM &   SEST 98 &  SEST 98 \\
220 & 16 30  3.69  &-39  9  0.1 &      DC 342.5+6.4  &  Har&   SEST 98 &  SEST 98 \\
221 & 16 31 18.69  &-23 41 21.2 &    CB65     L1704  &  Cle&   SEST 97 &  SEST 97 \\
222 & 16 31 35.63  &-24  1 27.1 &  L1709B(DCO+ 2-1)  &  Lor&   SEST 97 &  SEST 97 \\
223 & 16 32  2.34  &-24 56 11.4 &  L1689S(DCO+ 2-1)  &  Lor&   SEST 97 &  SEST 97 \\
224 & 16 32 22.66  &-24 28 33.0 &         IRAS16293  &  ???&   SEST 97 &  SEST 97 \\
225 & 16 32 42.03  &-23 54  8.6 &  L1709A(DCO+ 2-1)  &  Lor&   SEST 97 &  SEST 97 \\
226 & 16 33 26.60  &-36 36 34.3 &      DC 344.8+7.6  &  Har&   SEST 97 &  SEST 97 \\
227 & 16 34 33.17  &-15 47 10.8 &               L43  &  BM &   SEST 97 &  SEST 97 \\
228 & 16 34 37.47  &-15 47  0.5 &              L43E  &  BM &   SEST 97 &  SEST 97 \\
229 & 16 36 40.07  &-46  1 15.8 &      DC 338.2+0.8  &  Har&   SEST 97 & SEST 97  \\
230 & 16 36 50.22  &-35 37  8.4 &      DC 346.0+7.8  &  Har&   SEST 97 &  SEST 97 \\
231 & 16 36 51.75  &-35 56 32.3 &      DC 345.8+7.6  &  Har&   SEST 97 &  SEST 97 \\
232 & 16 37 16.11  &-35 31 36.7 &      DC 346.1+7.8  &  Har&   SEST 97 &  SEST 97 \\
\end{tabular}
\end{flushleft}
\end{table*}
\clearpage
\newpage

\begin{table*}
\begin{flushleft}
\begin{tabular}{rllcccc}
Nr  & RA (J2000) & Dec (J2000)       & Name               & Cat & Observatory & Observatory    \\
     & h  m  s   & \adeg\ \amin\ \asec   &                          &        & CS (2-1)       & CS (3-2)           \\  
\hline
233 & 16 37 27.64  &-35 13 53.8 &      DC 346.4+7.9  &  Har&   SEST 97 &  SEST 97 \\
234 & 16 37 46.00  &-35 26 22.6 &      DC 346.3+7.8  &  Har&   SEST 97 &  SEST 97 \\
235 & 16 38 52.73  &-35 49 48.1 &      DC 346.1+7.3  &  Har&   SEST 97 &  SEST 97 \\
236 & 16 39 30.35  &-14  5 50.5 &66     L121,MBM142  &  Cle&   SEST 97 &  SEST 97 \\
237 & 16 44 18.68  &-49 14 38.7 &      DC 336.7-2.3  &  Har&   SEST 97 &  SEST 97 \\
238 & 16 45  1.03  &-47 45 11.7 &      DC 337.9-1.4  &  Har&   SEST 97 &  SEST 97 \\
239 & 16 44 50.26  &-40 23  5.9 &      DC 343.4+3.5  &  Har&   SEST 97 &  SEST 97 \\
240 & 16 46 45.13  &-46 27 22.4 &      DC 339.1-0.8  &  Har&   SEST 98 &  SEST 98 \\
241 & 16 46 44.78  &-44 30 46.3 &      DC 340.5+0.5  &  Har&   SEST 97 &  SEST 97 \\
242 & 16 47  6.64  & -9 35 21.0 &              L260  &  BM &   SEST 97 &  SEST 97 \\
243 & 16 47 23.12  &-13 59 21.1 &              L158  &  BM &   SEST 97 &  SEST 97 \\
244 & 16 48  6.81  &-10 51 47.9 &             L234A  &  BM &   SEST 97 &  SEST 97 \\
245 & 16 48  8.50  &-10 56 57.8 &             L234E  &  BM &   SEST 97 &  SEST 97 \\
246 & 16 49  9.40  &-39 45  6.0 &      DC 344.5+3.2  &  Har&   SEST 97 &  SEST 97 \\
247 & 16 50 15.42  &-18  6  6.3 &               L63  &  BM &   SEST 97 &  SEST 97 \\
248 & 16 50 33.70  &-19  7 50.1 &      CB67     L31  &  Cle&   SEST 98 &  SEST 98 \\
249 & 16 54 32.26  &-40 31  7.6 &      DC 344.5+2.0  &  Har&   SEST 97 &  SEST 97 \\
250 & 16 57 16.24  &-16  9 21.9 &  CB68     L146(?)  &  Cle&   SEST 97 &  SEST 97 \\
251 & 16 58 41.82  &-50 35 45.0 &      DC 337.1-4.9  &  Har&   SEST 96 &  SEST 96 \\
252 & 16 58 25.54  &-36 42 15.1 &      DC 348.0+3.7  &  Har&   SEST 97 &  SEST 97 \\
253 & 17  2 20.01  &-33 16 34.4 &      DC 351.2+5.2  &  Har&   SEST 97 &  SEST 97 \\
254 & 17  2 42.04  &-33 16 59.8 &      CB69     B49  &  Cle&   SEST 97 &  SEST 97 \\
255 & 17  3  0.05  &-22 13 39.0 &       CB70     L4  &  Cle&   SEST 97 & SEST 97  \\
256 & 17  4 26.20  &-36 18 49.7 &      DC 349.0+3.0  &  Har&   SEST 97 &  SEST 97 \\
257 & 17  4 26.90  &-36  8 25.6 &      DC 349.2+3.1  &  Har&   SEST 97 &  SEST 97 \\
258 & 17  4 45.97  &-36 10 18.3 &      DC 349.2+3.1  &  Har&   SEST 97 &  SEST 97 \\
259 & 17  4 48.78  &-36  3 36.1 &      DC 349.3+3.1  &  Har&   SEST 97 &  SEST 97 \\
260 & 17  5 10.11  &-22 12 55.8 &              CB71  &  Cle&   SEST 97 &  SEST 97 \\
261 & 17  6 39.18  &-42 33 22.7 &      DC 344.3-1.1  &  Har&   SEST 97 &  SEST 97 \\
262 & 17  6 39.79  &-34 13 52.1 &      DC 351.0+3.9  &  Har&   SEST 97 &  SEST 97 \\
263 & 17  8  2.31  &-32  2 52.1 &      DC 352.9+5.0  &  Har&   SEST 97 &  SEST 97 \\
264 & 17  8 21.75  &-36 30 51.0 &      DC 349.3+2.3  &  Har&   SEST 97 &  SEST 97 \\
265 & 17  8 30.11  &-22 53 49.7 &  CB72     B57,L11  &  Cle&   SEST 97 &  SEST 97 \\
266 & 17  8 47.45  &-32  6 53.9 &CB73     B56,L1685  &  Cle&   SEST 97 &  SEST 97 \\
267 & 17 10 44.33  &-40 31 35.2 &      DC 346.4-0.5  &  Har&   SEST 97 &  SEST 97 \\
268 & 17 11 48.64  &-22 28 35.6 &B74     B60,L38(?)  &  Cle&   SEST 97 &  SEST 97 \\
269 & 17 13  3.75  &-30 15 42.6 &     CB75     B247  &  Cle&   SEST 97 &  SEST 97 \\
270 & 17 13 29.13  &-34 15 23.0 &      DC 351.8+2.8  &  Har&   SEST 97 &  SEST 97 \\
271 & 17 15  2.11  &-20 29 59.7 & CB76     B61,L111  &  Cle&   SEST 98 &  SEST 98 \\
272 & 17 15 30.21  &-32 24 56.3 &      DC 353.5+3.5  &  Har&   SEST 98 &  SEST 98 \\
273 & 17 15 14.63  & -1 43 20.0 &              CB77  &  Cle&   SEST 98 &  SEST 98 \\
274 & 17 19 35.85  &-43 27  3.5 &      DC 345.0-3.5  &  Har&   SEST 96 & SEST 96  \\
275 & 17 19 43.72  &-26 43  1.9 &CB79     B65,L1772  &  Cle&   SEST 97 &  SEST 97 \\
276 & 17 20 56.20  &-33 31 21.0 &      DC 353.3+1.9  &  Har&   SEST 97 &  SEST 97 \\
277 & 17 21 47.49  &-44  8 48.1 &      DC 344.6-4.3  &  Har&   SEST 96 &  SEST 96 \\
278 & 17 21 49.98  &-21 48 23.6 &     CB80     L101  &  Cle&  SEST 97  &  SEST 97 \\
279 & 17 22 37.33  &-27  5  1.4 &    CB81     L1774  &  Cle&   SEST 97 &  SEST 97 \\
280 & 17 22 36.75  &-23 50 34.3 &  CB82     B68,L55  &  Cle&   SEST 97 &  SEST 97 \\
281 & 17 25 11.53  &-40 16 27.1 &      DC 348.2-2.6  &  Har&   SEST 96 &  SEST 96 \\
282 & 17 30 37.67  &-43  8 27.9 &      DC 346.4-5.0  &  Har&   SEST 96 &  SEST 96 \\
283 & 17 32 14.16  &-25 25 28.8 &      CB89     L52  &  Cle&   SEST 96 &  SEST 96 \\
284 & 17 35 10.66  &-33 32 37.5 &      DC 354.9-0.6  &  Har&   SEST 96 &  SEST 96 \\
285 & 17 35 26.36  &-33 21 36.3 &      DC 355.1-0.5  &  Har&   SEST 96 &  SEST 96 \\
286 & 17 38 26.52  &-19 44 30.5 &  CB90     L216(?)  &  Cle&   SEST 96 &  SEST 96 \\
287 & 17 39 33.36  &-19 36 35.7 &     CB91     L219  &  Cle&   SEST 96 &  SEST 96 \\
288 & 17 40 44.44  &-19 39 51.5 &     CB92     L226  &  Cle&   SEST 96 &  SEST 96 \\
289 & 17 40 49.54  &-19 44 20.2 &     CB93     L223  &  Cle&   SEST 96 &  SEST 96 \\
290 & 17 41 22.61  &-19 47 38.8 &     CB94     L222  &  Cle&  SEST 96  &  SEST 96 \\
\end{tabular}
\end{flushleft}
\end{table*}
\clearpage
\newpage

\begin{table*}
\begin{flushleft}
\begin{tabular}{rllcccc}
Nr  & RA (J2000) & Dec (J2000)       & Name               & Cat & Observatory & Observatory    \\
     & h  m  s   & \adeg\ \amin\ \asec   &                          &        & CS (2-1)       & CS (3-2)           \\  
\hline
291 & 17 45 20.94  &-20  0 30.5 &CB95     B83, L233  &  Cle&   SEST 96 &  SEST 96 \\
292 & 17 46 23.07  &-20  6  2.0 &     CB96     L235  &  Cle&   SEST 96 &  SEST 96 \\
293 & 17 46 47.31  &-20 16 37.2 & CB97     B84,L235  &  Cle&   SEST 96 &  SEST 96 \\
294 & 17 47  0.62  &-20 30 29.3 &              CB98  &  Cle&   SEST 96 &  SEST 96 \\
295 & 17 47  3.34  &-20 17 43.1 &     CB99     L235  &  Cle&   SEST 96 &  SEST 96 \\
296 & 17 51 45.16  & -2 58 48.8 &             CB100  &  Cle&   SEST 96 &  SEST 96 \\
297 & 17 53 16.56  & -8 25  1.4 &     CB101    L392  &  Cle&   SEST 96 &  SEST 96 \\
298 & 17 54 42.93  &-27 49 56.0 &             CB102  &  Cle&   SEST 96 &  SEST 96 \\
299 & 17 55  6.42  &-32 37  4.6 &      DC 357.9-3.6  &  Har&   SEST 96 &  SEST 96 \\
300 & 17 55 12.17  &-34 15 40.2 &      DC 356.5-4.5  &  Har&   SEST 96 &  SEST 96 \\
301 & 17 55 24.96  & -2 48 44.8 &             CB103  &  Cle&   SEST 96 &  SEST 96 \\
302 & 17 55 31.38  & -8 15 25.5 &     CB104    L400  &  Cle&   SEST 96 &  SEST 96 \\
303 & 17 58  6.09  & -3 46 21.1 &     CB105    L460  &  Cle&   SEST 96 &  SEST 96 \\
304 & 18  1  9.78  &-32 49 44.1 &      DC 358.3-4.8  &  Har&   SEST 96 &  SEST 96 \\
305 & 18  0 34.79  & -3 30 36.2 &     CB106    L468  &  Cle&   SEST 96 &  SEST 96 \\
306 & 18  1 17.02  &-32 58 13.6 &      DC 358.2-4.9  &  Har&   SEST 96 &  SEST 96 \\
307 & 18  3  2.48  &-46 39 48.9 &     DC 346.2-11.7  &  Har&   SEST 96 &  SEST 96 \\
308 & 18  2 58.01  &-27 52  7.9 &  CB107    B86,L93  &  Cle&   SEST 96 &  SEST 96 \\
309 & 18  3 15.10  &-33  1  5.0 &      DC 358.4-5.3  &  Har&   SEST 96 &  SEST 96 \\
310 & 18  3  4.13  &-20 51  0.1 &     CB108    L262  &  Cle&   SEST 96 &  SEST 96 \\
311 & 18  5  3.85  &-28 25 53.8 &  CB109    SCHO741  &  Cle&   SEST 96 &  SEST 96 \\
312 & 18  5 55.94  &-18 25 13.5 &     CB110    L307  &  Cle&   SEST 96 &  SEST 96 \\
313 & 18  7 13.85  &-18 21 14.8 &     CB111    L310  &  Cle&   SEST 96 &  SEST 96 \\
314 & 18  8 35.83  & -1 50  1.1 &  CB112    L502(?)  &  Cle&   SEST 96 &  SEST 96 \\
315 & 18 10 24.67  &-22 44 23.1 &     CB113    L249  &  Cle&   SEST 96 &  SEST 96 \\
316 & 18 12 13.57  &-22 40 20.2 &             CB114  &  Cle&   SEST 96 &  SEST 96 \\
317 & 18 11 59.95  & -7 54  0.5 &     CB115    L426  &  Cle&   SEST 96 &  SEST 96 \\
318 & 18 12 20.68  &-22 45 21.7 &             CB116  &  Cle&   SEST 96 &  SEST 96 \\
319 & 18 12  8.53  & -7 32 42.9 &     CB117    L430  &  Cle&   SEST 96 &  SEST 96 \\
320 & 18 12 35.61  &-15 49 11.2 &             CB118  &  Cle&   SEST 96 &  SEST 96 \\
321 & 18 13 52.98  & -7  4 36.2 &     CB119    L438  &  Cle&   SEST 96 &  SEST 96 \\
322 & 18 14  5.05  & -7  8 30.4 &     CB120    L438  &  Cle&   SEST 96 &  SEST 96 \\
323 & 18 14  9.85  & -6 58 26.0 &             CB121  &  Cle&   SEST 96 &  SEST 96 \\
324 & 18 14 28.04  & -7  7 54.7 &     CB122    L438  &  Cle&   SEST 96 &  SEST 96 \\
325 & 18 14 37.34  & -7 23  2.0 &     CB123    L436  &  Cle&   SEST 96 &  SEST 96 \\
326 & 18 14 43.42  &  7  4 38.0 &      CB124 LBN102  &  Cle&   SEST 97 &  SEST 97 \\
327 & 18 15 34.60  &-18 11 12.3 & CB125    B92,L323  &  Cle&   SEST 96 &  SEST 96 \\
328 & 18 15 17.08  & -3 45 37.0 &  CB126    L490(?)  &  Cle&   SEST 96 &  SEST 96 \\
329 & 18 15 33.37  &-16 26 10.3 &             CB127  &  Cle&   SEST 96 &  SEST 96 \\
330 & 18 15 49.19  & -3 51 10.6 &     CB128    L492  &  Cle&   SEST 96 &  SEST 96 \\
331 & 18 15 49.06  & -3 44 27.6 &     CB129    L492  &  Cle&   SEST 96 &  SEST 96 \\
332 & 18 16 14.67  & -2 32 46.7 &  CB130    L507(?)  &  Cle&   SEST 96 &  SEST 96 \\
333 & 18 16 59.41  &-18  2 44.2 & CB131    B93,L328  &  Cle&   SEST 96 &  SEST 96 \\
334 & 18 19 46.82  & -6  5 51.5 &     CB132    L466  &  Cle&   SEST 96 &  SEST 96 \\
335 & 18 22 29.40  & -1 27 32.5 &             CB133  &  Cle&   SEST 96 &  SEST 96 \\
336 & 18 22 44.69  & -1 42 40.4 &             CB134  &  Cle&   SEST 96 &  SEST 96 \\
337 & 18 23 31.91  &-20 45 37.8 &     CB135    L306  &  Cle&   SEST 96 &  SEST 96 \\
338 & 18 23 46.20  & -1 17 30.9 &     CB136    L533  &  Cle&   SEST 96 &  SEST 96 \\
339 & 18 24 18.85  &  0 58 52.5 &     CB137    L539  &  Cle&   SEST 96 &  SEST 96 \\
340 & 18 24 57.48  &-10 53  0.1 &     CB138    L414  &  Cle&   SEST 96 &  SEST 96 \\
341 & 18 25 30.21  &-10 39 54.7 &139    B94,L416,L4  &  Cle&   SEST 96 &  SEST 96 \\
342 & 18 26 14.78  &-10 18 11.5 &140    B96,L420(?)  &  Cle&   SEST 96 &  SEST 96 \\
343 & 18 27 51.16  &-11 27 30.6 &     CB141    L411  &  Cle&   SEST 96 &  SEST 96 \\
344 & 18 29 53.85  &-13 41 14.8 &             CB142  &  Cle&   SEST 96 &  SEST 96 \\
345 & 18 29 57.00  &  1 13 15.1 &         Serp SMM4  & LLM &   OSO  98 &  KITT 97 \\
346 & 18 30 28.88  &-17 42 53.4 &CB143    B311,L356  &  Cle&   SEST 96 &  SEST 96 \\
347 & 18 30 40.82  &-16  5 20.5 &  CB144    L374(?)  &  Cle&   SEST 96 &  SEST 96 \\
348 & 18 32 24.39  & -9 10 29.7 &  CB145    B1,L443  &  Cle&  SEST 96  &  SEST 96 \\
\end{tabular}
\end{flushleft}
\end{table*}
\clearpage
\newpage

\begin{table*}
\begin{flushleft}
\begin{tabular}{rllcccc}
Nr  & RA (J2000) & Dec (J2000)       & Name               & Cat & Observatory & Observatory    \\
     & h  m  s   & \adeg\ \amin\ \asec   &                          &        & CS (2-1)       & CS (3-2)           \\  
\hline
349 & 18 32 32.15  & -8 57 44.2 &     CB146    B101  &  Cle&   SEST 96 &  SEST 96 \\
350 & 18 32 52.46  & -9 13 49.7 &  CB147    B1,L443  &  Cle&   SEST 96 &  SEST 96 \\
351 & 18 33 18.01  &-26  1 33.6 & CB148    B98,L239  &  Cle&   SEST 96 &  SEST 96 \\
352 & 18 38  8.83  &-13 43 58.1 &CB149    B102,L401  &  Cle&   SEST 96 &  SEST 96 \\
353 & 18 38 55.03  & 13 23 24.4 &            CB150   &  Cle&   SEST 97 &  SEST 97 \\
354 & 18 39 20.32  & 13  8 53.2 &            CB151   &  Cle&   SEST 97 &  SEST 97 \\
355 & 18 41 36.19  & -2  9 16.7 &CB152    B316,L555  &  Cle&   SEST 96 &  SEST 96 \\
356 & 18 42 28.63  &-19 59 13.7 &CB153    B315,L346  &  Cle&   SEST 96 &  SEST 96 \\
357 & 18 47 18.94  & -4 33  6.3 &     CB155    B104  &  Cle&   SEST 96 &  SEST 96 \\
358 & 18 48 56.54  & -5  4 52.3 &     CB156    B106  &  Cle&  SEST 96  &  SEST 96 \\
359 & 18 53  2.70  & -6 59  0.9 &CB157    B114,L514  &  Cle&   SEST 96 &  SEST 96 \\
360 & 18 53 20.32  & -6 39 31.6 &CB158    B115,L518  &  Cle&   SEST 96 &  SEST 96 \\
361 & 18 53 22.45  & -6 46 14.5 &  CB159    SCHO943  &  Cle&   SEST 96 &  SEST 96 \\
362 & 18 53 40.20  & -7 25 30.2 &     CB160    B117  &  Cle&  SEST 96  &  SEST 96 \\
363 & 18 53 56.22  & -7 26 29.1 &     CB161    B118  &  Cle&   SEST 96 &  SEST 96 \\
364 & 18 54 23.97  & -7 13 51.1 &  CB162    SCHO950  &  Cle&   SEST 96 &  SEST 96 \\
365 & 18 54 34.90  & -4 33 17.2 &     CB163    B119  &  Cle&   SEST 96 &  SEST 96 \\
366 & 18 54 50.95  & -4 36  3.1 &     CB164    B120  &  Cle&   SEST 96 &  SEST 96 \\
367 & 18 55 44.77  & -4 26 29.3 &     CB165    B322  &  Cle&   SEST 96 &  SEST 96 \\
368 & 18 56 47.12  & -4 45 25.9 &CB166    B122,L545  &  Cle&   SEST 96 &  SEST 96 \\
369 & 18 57 32.10  & -4 44 15.7 &CB167    B123,L546  &  Cle&   SEST 96 &  SEST 96 \\
370 & 18 58 59.87  & -4 32 30.5 &CB168    B126,L556  &  Cle&   SEST 96 &  SEST 96 \\
371 & 18 59 24.83  & -4 30 48.7 &CB169    B126,L556  &  Cle&   SEST 96 &  SEST 96 \\
372 & 19  1 51.33  &-36 58 51.9 &     DC 359.9-17.9  &  Har&   SEST 96 &  SEST 96 \\
373 & 19  1 31.89  & -5 27 19.8 &CB170    B127,L544  &  Cle&   SEST 96 &  SEST 96 \\
374 & 19  1 40.88  & -4 33 58.1 &     CB171    B128  &  Cle&   SEST 96 &  SEST 96 \\
375 & 19  1 44.78  & -5 21 43.9 &             CB172  &  Cle&   SEST 96 &  SEST 96 \\
376 & 19  1 45.99  & -5 32 50.8 &     CB173    B130  &  Cle&   SEST 96 &  SEST 96 \\
377 & 19  1 58.01  & -5 33 57.0 &     CB174    B130  &  Cle&   SEST 96 &  SEST 96 \\
378 & 19  2  5.70  & -5 17 51.4 &CB175    B129,L549  &  Cle&   SEST 96 &  SEST 96 \\
379 & 19  2 13.67  & -4 22 44.8 &             CB176  &  Cle&   SEST 96 &  SEST 96 \\
380 & 19  4 36.75  & -5 21  1.8 &             CB179  &  Cle&   SEST 96 &  SEST 96 \\
381 & 19 10  8.42  &-78 36  2.5 &     DC 315.8-27.5  &  Har&   SEST 96 &  SEST 96 \\
382 & 19 10 57.12  &-78 36  5.1 &     DC 315.8-27.5  &  Har&   SEST 96 &  SEST 96 \\
383 & 19  6 12.48  & -6 53 27.2 &CB180    B133,L531  &  Cle&   SEST 97 &  SEST 97 \\
384 & 19 12 13.12  & 17 57 50.9 &        CB182 L714  &  Cle&   SEST 97 &  SEST 97 \\
385 & 19 13 20.83  & 16 34 38.6 &            CB183   &  Cle&   SEST 97 &  SEST 97 \\
386 & 19 13 52.99  & 16 27 24.8 &        CB184 L709  &  Cle&   OSO  98 &  KITT 97 \\
387 & 19 15 56.19  & -1 19  7.4 &     CB185    B137  &  Cle&   SEST 97 &  SEST 97 \\
388 & 19 16  7.14  & -1 16 18.6 &     CB186    B137  &  Cle&   SEST 97 &  SEST 97 \\
389 & 19 20 16.83  & 11 36 12.0 &            CB188   &  Cle&   OSO  98 &  KITT 97 \\
390 & 19 20 30.94  & 11 30 37.0 &        CB189 L676  &  Cle&   SEST 97 &  SEST 97 \\
391 & 19 20 47.46  & 23 29 33.6 &        CB190 L771  &  Cle&   OSO  98 &  KITT 97 \\
392 & 19 23 19.95  & 12 25 51.6 &        CB192 L686  &  Cle&   SEST 97 &  SEST 97 \\
393 & 19 23 51.46  & 11  7  0.7 &        CB193 L675  &  Cle&   SEST 97 &  SEST 97 \\
394 & 19 26 32.48  & 23 58 42.2 &              L778  &  BM &   OSO  98 &  KITT 97 \\
395 & 19 29 55.67  & 14 34 39.5 &            CB194   &  Cle&   SEST 97 &  SEST 97 \\
396 & 19 34 47.44  & 12 14 30.0 &        CB195 L701  &  Cle&   SEST 97 &  SEST 97 \\
397 & 19 35 15.36  & 12 19 46.9 &        CB196 B334  &  Cle&   SEST 97 &  SEST 97 \\
398 & 19 36  6.42  & 18 23 35.5 &            CB197   &  Cle&   SEST 97 &  SEST 97 \\
399 & 19 36 35.40  & 12 19 39.2 &   CB198 B336,L702  &  Cle&   SEST 98 &       no \\
400 & 19 37  1.03  &  7 34 10.8 &   CB199 B335,L663  &  Cle&    OSO 98 &  KITT 97 \\
401 & 19 41 30.73  & 11 13 59.8 &     CB200 B143,L7  &  Cle&   SEST 97 &  SEST 97 \\
402 & 19 42 10.60  & 11 22  4.4 &    CB201 SCHO1080  &  Cle&   SEST 97 &  SEST 97 \\
403 & 19 42 31.13  & 18 52  9.1 &            CB202   &  Cle&   OSO  98 &  KITT 97 \\
404 & 19 43 55.94  & 19  4 59.6 &  CB203 SCHO1091(?)  &  Cle&   SEST 97 &  SEST 97 \\
405 & 19 44 40.63  &  8 37 48.2 &        CB204 L680  &  Cle&   SEST 98 &       no \\
406 & 19 45 27.05  & 27 52 21.0 &        CB205 L810  &  Cle&    OSO 97 &  KITT 97 \\
\end{tabular}
\end{flushleft}
\end{table*}
\clearpage
\newpage

\begin{table*}
\begin{flushleft}
\begin{tabular}{rllcccc}
Nr  & RA (J2000) & Dec (J2000)       & Name               & Cat & Observatory & Observatory    \\
     & h  m  s   & \adeg\ \amin\ \asec   &                          &        & CS (2-1)       & CS (3-2)           \\  
\hline
407 & 19 46 30.51  & 21 13 31.8 &            CB207   &  Cle&   OSO  98 &  KITT 97 \\
408 & 19 54 40.48  & 33 47 23.0 &        CB210 L832  &  Cle&    OSO 97 &       no \\
409 & 19 59 46.80  & 24 55 29.2 &        CB211 L802  &  Cle&   OSO  98 &  KITT 97 \\
410 & 20  1 31.18  & 24 42 46.7 &        CB213 L805  &  Cle&   OSO  98 &  KITT 97 \\
411 & 20  3 59.96  & 26 39 17.2 &      CB214 L814(?)  &  Cle&   OSO  98 &  KITT 97 \\
412 & 20  4 14.83  & 26 46  1.1 &      CB215 L814(?)  &  Cle&   OSO  98 &  KITT 97 \\
413 & 20  7  6.66  & 27 28 52.8 &        IRAS 20050  &  ???&   OSO  98 &  KITT 97 \\
414 & 20  7 44.82  & 37  5 53.6 &        CB217 L863  &  Cle&    OSO 97 &       no \\
415 & 20 18 18.80  & 63 52 39.4 &      CB219 LBN444  &  Cle&    OSO 97 &       no \\
416 & 20 23 19.91  & 68  1 55.0 &            CB220   &  Cle&    OSO 97 &  KITT 97 \\
417 & 20 31 41.99  & 25 59 51.2 &            CB221   &  Cle&    OSO 97 &       no \\
418 & 20 33 30.47  & 64  1 27.8 &       CB222 L1094  &  Cle&    OSO 97 &  KITT 97 \\
419 & 20 34 46.20  & 64 10 45.1 &            CB223   &  Cle&    OSO 97 &  KITT 97 \\
420 & 20 35 50.18  & 67 54 22.3 &             L1152  &  BM &    OSO 97 &  KITT 97 \\
421 & 20 36 17.14  & 63 53 15.2 &       CB224 L11(?)  &  Cle&    OSO 97 &  KITT 97 \\
422 & 20 37 51.42  & 56 18 19.7 &            CB225   &  Cle&    OSO 97 &  KITT 97 \\
423 & 20 43 29.88  & 67 52 41.9 &            L1155C  &  BM &    OSO 97 &       no \\
424 & 20 48 18.17  & 59 38 21.6 &        CB226 B148  &  Cle&    OSO 97 &  KITT 97 \\
425 & 20 49 18.90  & 59 31 12.9 &        CB227 B149  &  Cle&    OSO 97 &       no \\
426 & 20 50 43.14  & 44 21 59.4 &   Sug31 HII(S117)  &  S91&    OSO 97 &  KITT 97 \\
427 & 20 51 20.41  & 56 15 45.1 &            CB228   &  Cle&    OSO 97 &  KITT 97 \\
428 & 20 51 27.42  & 60 18 59.8 &            L1082C  &  BM &    OSO 97 & KITT 97  \\
429 & 20 53 29.57  & 60 14 40.3 &            L1082A  &  BM &    OSO 97 &  KITT 97 \\
430 & 20 53 50.28  & 60  9 46.4 &            L1082B  &  BM &    OSO 97 &  KITT 97 \\
431 & 20 53 32.09  & 68 18 27.4 &       CB229 L1171  &  Cle&    OSO 97 &  KITT 97 \\
432 & 21  0 22.43  & 68 12 52.8 &             L1174  &  BM &    OSO 97 &  KITT 97 \\
433 & 21  2 23.97  & 67 54  7.9 &             L1172  &  BM &    OSO 97 &  KITT 97 \\
434 & 21 12 20.11  & 47 24 24.1 &              B361  &  BM &   OSO  98 &  KITT 97 \\
435 & 21 17 40.83  & 68 18 23.0 &       CB230 L1177  &  Cle&    OSO 97 &  KITT 97 \\
436 & 21 32 23.66  & 57 24  8.3 &   Sug32 HII(S131)  &  S91&   OSO  98 &  KITT 97 \\
437 & 21 33 12.13  & 57 29 34.4 &   Sug33 HII(S131)  &  S91&   OSO  98 &  KITT 97 \\
438 & 21 33 32.28  & 58  3 28.3 &   Sug34 HII(S131)  &  S91&    OSO 97 &  KITT 97 \\
439 & 21 33 55.53  & 54 41 29.2 &        CB231 B157  &  Cle&    OSO 97 &       no \\
440 & 21 36  5.38  & 58 31 39.0 &   Sug35 HII(S131)  &  S91&   OSO  98 &  KITT 97 \\
441 & 21 36 12.31  & 57 27 34.2 &   Sug36 HII(S131)  &  S91&   OSO  98 &  KITT 97 \\
442 & 21 37 13.82  & 43 21 34.3 &        CB232 B158  &  Cle&   OSO  98 &  KITT 97 \\
443 & 21 40 28.92  & 56 35 58.0 &   Sug37 HII(S131)  &  S91&   OSO  98 &  KITT 97 \\
444 & 21 40 33.03  & 57 48  6.2 &        CB233 B161  &  Cle&    OSO 97 &  KITT 97 \\
445 & 21 40 42.27  & 58 16  9.7 &   Sug38 HII(S131)  &  S91&   OSO  98 &  KITT 97 \\
446 & 21 40 33.40  & 70 18 33.2 &      CB234 LBN515  &  Cle&    OSO 97 &  KITT 97 \\
447 & 21 46  6.75  & 57 26 22.9 &   Sug39 HII(S131)  &  S91&   OSO  98 &  KITT 97 \\
448 & 21 46 14.62  & 57  8 59.2 &   Sug40 HII(S131)  &  S91&   OSO  98 &  KITT 97 \\
449 & 21 46 29.17  & 57 18 40.8 &   Sug41 HII(S131)  &  S91&   OSO  98 &  KITT 97 \\
450 & 21 46 36.74  & 57 12 25.1 &   Sug42 HII(S131)  &  S91&   OSO  98 &  KITT 97 \\
451 & 21 47 25.69  & 47 32  9.7 &            L1031B  &  BM &   OSO  98 &  KITT 97 \\
452 & 21 56 18.45  & 59  1  9.8 &       CB235 L1142  &  Cle&   OSO  98 &  KITT 97 \\
453 & 22  5 25.10  & 59 33 32.6 &       CB236 L1166  &  Cle&    OSO 97 &  KITT 97 \\
454 & 22  7 21.70  & 59 40 14.7 &  CB237 B173,L1169  &  Cle&    OSO 97 &  KITT 97 \\
455 & 22 13 22.48  & 41  1 31.4 &            CB238   &  Cle&    OSO 97 &  KITT 97 \\
456 & 22  9 56.65  & 86 44  2.7 &            CB239   &  Cle&    OSO 97 &       no \\
457 & 22 28 52.15  & 64 13 43.4 &   Sug44 HII(S145)  &  S91&    OSO 97 &  KITT 97 \\
458 & 22 30 31.70  & 75 14 16.8 &       L1251A(2,0)  &  BM &    OSO 97 &  KITT 97 \\
459 & 22 31 18.79  & 75 13 18.1 &      L1251A(5,-1)  &  BM &    OSO 97 &  KITT 97 \\
460 & 22 33 48.34  & 58 33 30.5 &       CB240 L1192  &  Cle&    OSO 97 &  KITT 97 \\
461 & 22 47 48.74  & 58  2 50.4 &   Sug43 HII(S142)  &  S91&    OSO 97 &       no \\
462 & 23 11 41.78  & 66  3  6.9 &       CB241 L1239  &  Cle&   OSO  98 & KITT 97  \\
463 & 23 11 57.51  & 61 39  4.1 &       CB242 L1225  &  Cle&   OSO  98 &  KITT 97 \\
464 & 23 25 12.82  & 63 36 29.8 &       CB243 L1246  &  Cle&   OSO  98 &  KITT 97 \\
\end{tabular}
\end{flushleft}
\end{table*}
\clearpage
\newpage

\begin{table*}
\begin{flushleft}
\begin{tabular}{rllcccc}
Nr  & RA (J2000) & Dec (J2000)       & Name               & Cat & Observatory & Observatory    \\
     & h  m  s   & \adeg\ \amin\ \asec   &                          &        & CS (2-1)       & CS (3-2)           \\  
\hline
465 & 23 25 29.56  & 74 18 15.0 &            L1262A  &  BM &    OSO 97 &  KITT 97 \\
466 & 23 25 48.71  & 74 17 37.2 & CB244 L1262,L1261  &  Cle&    OSO 97 &  KITT 97 \\
467 & 23 55 21.86  & 58 51 21.7 &    CB245 SCHO1448  &  Cle&    OSO 97 &       no \\
468 & 23 56 43.54  & 58 34 28.8 &       CB246 L1253  &  Cle&    OSO 97 &  KITT 97 \\
469 & 23 57 25.32  & 64 49 26.9 &       CB247 L1263  &  Cle&    OSO 97 &       no \\
470 & 23 59 25.71  & 67 23 39.0 &    Sug1 HII(S171)  &  S91&    OSO 97 &  KITT 97 \\
471 &  0  1 49.86  & 48  6  8.0 &            CB248   &  Cle&   OSO  98 & KITT 97  \\
\end{tabular}
\end{flushleft}
 {\bf Catalogue id} \\
\\
BM - Benson \& Myers (1989) \\
Cle  - Clemens \& Barvainis (1988)\\
Har - Hartley et al. (1996) \\
LLM - Larsson et al. (2000) \\
Lor -   Loren et al. (1990) \\
S91 - Sugitani et al. (1991) 
\end{table*}


\end{document}